\documentclass[12pt]{article}
\usepackage{hyperref}
\usepackage[numbers,compress]{natbib}
\usepackage{graphicx,psfrag,epsf,color}
\usepackage[nointlimits]{amsmath}
\usepackage{amssymb,calc,graphicx}

\newcommand{\braket}[1]{\langle #1 \rangle}
\newcommand{\Tprod}[1]{{\mathrm T}\lbrack #1 \rbrack}

\newcommand{\Li}[1]{{\mathrm{Li}_{#1}}}

\newcommand{\Oc}{\mathcal{O}}

\newcommand{\eps}{\epsilon}
\newcommand{\ep}{\text{ep}}
\newcommand{\IR}{\text{IR}}
\newcommand{\UV}{\text{UV}}
\newcommand{\Nc}{N_{\text{c}}}
\newcommand{\CF}{C_{\text{F}}}
\newcommand{\tL}{\text{L}}
\newcommand{\tR}{\text{R}}
\newcommand{\tw}{w}
\newcommand{\sw}{s_\tw}
\newcommand{\cw}{c_\tw}

\newlength{\dslashwidth}
\newcommand{\dslash}[2][0.55]{%
  \settowidth{\dslashwidth}{#2}%
  \hspace*{#1\dslashwidth}\makebox[0pt]{/}\hspace*{-#1\dslashwidth}%
  #2%
}

\def\lsim{\mathrel{\raise.3ex\hbox{$<$\kern-.75em\lower1ex\hbox{$\sim$}}}}
\def\gsim{\mathrel{\raise.3ex\hbox{$>$\kern-.75em\lower1ex\hbox{$\sim$}}}}

\catcode`@=11
\newcount\@tempcntc
\def\@citex[#1]#2{\if@filesw\immediate\write\@auxout{\string\citation{#2}}\fi
  \@tempcnta\z@\@tempcntb\m@ne\def\@citea{}\@cite{\@for\@citeb:=#2\do
    {\@ifundefined
       {b@\@citeb}{\@citeo\@tempcntb\m@ne\@citea\def\@citea{,}{\bf ?}\@warning
       {Citation `\@citeb' on page \thepage \space undefined}}%
    {\setbox\z@\hbox{\global\@tempcntc0\csname b@\@citeb\endcsname\relax}%
     \ifnum\@tempcntc=\z@ \@citeo\@tempcntb\m@ne
       \@citea\def\@citea{,}\hbox{\csname b@\@citeb\endcsname}%
     \else
      \advance\@tempcntb\@ne
      \ifnum\@tempcntb=\@tempcntc
      \else\advance\@tempcntb\m@ne\@citeo
      \@tempcnta\@tempcntc\@tempcntb\@tempcntc\fi\fi}}\@citeo}{#1}}
\def\@citeo{\ifnum\@tempcnta>\@tempcntb\else\@citea\def\@citea{,}%
  \ifnum\@tempcnta=\@tempcntb\the\@tempcnta\else
   {\advance\@tempcnta\@ne\ifnum\@tempcnta=\@tempcntb \else \def\@citea{--}\fi
    \advance\@tempcnta\m@ne\the\@tempcnta\@citea\the\@tempcntb}\fi\fi}
\catcode`@=12

\begin{document}
\allowdisplaybreaks

\begin{titlepage}

\begin{flushright}
TTK-13-17\\
SFB/CPP-13-43 \\
IFIC/13-29\\[0.2cm]
\end{flushright}

\vskip1.5cm
\begin{center}
\Large\bf\boldmath
NNLO non-resonant corrections to threshold top-pair production from $e^+ e^-$ collisions: Endpoint-singular terms
\end{center}

\vspace{1cm}
\begin{center}
{\sc B.~Jantzen$^{a}$} and  {\sc P. Ruiz-Femen\'\i a$^{b}$}\\[5mm]
{\it ${}^a$Institut f\"ur Theoretische Teilchenphysik und Kosmologie,\\
RWTH Aachen University,\\
D-52056 Aachen, Germany}\\[0.3cm]
{\it ${}^b$Instituto de F\'\i sica Corpuscular (IFIC), 
CSIC-Universitat de Val\`encia \\
Apartado de Correos 22085, E-46071 Valencia, Spain}\\[0.3cm]
\end{center}

\vspace{2cm}
\begin{abstract}
\noindent
We analyse the subleading non-resonant contributions to the 
$e^+e^-\to W^+ W^- b\bar{b}$ cross section at energies near the 
top--antitop threshold. These correspond to next-to-next-to-leading-order
(NNLO) corrections with respect to the leading-order resonant result.
We show that these corrections produce $1/\epsilon$ endpoint singularities 
which precisely cancel the finite-width divergences arising in the resonant
production of the $W^+ W^- b\bar{b}$ final state from on-shell decays 
of the top and antitop quarks at the same order. We also provide analytic results
for the $(m_t/\Lambda)^2$, $(m_t/\Lambda)$ and $(m_t/\Lambda)^0\log \Lambda$
terms that dominate the expansion in powers of $(\Lambda/m_t)$ 
of the complete set of NNLO non-resonant corrections, where $\Lambda$ is a cut imposed on the 
invariant masses of the $bW$ pairs that is neither too tight nor too loose
($m_t \Gamma_t \ll \Lambda^2 \ll m_t^2$). 
\end{abstract}
\end{titlepage}

\section{Introduction}

An $e^+e^-$ linear collider (LC) is probably the most compelling
case as the next-generation particle collider for high-precision physics.
One of the LC options,
the International Linear Collider, has recently completed its technical design,
and there is hope that funding to begin its construction could be gathered up in a near future,
especially if Japan's proposal to host it goes forward. 
The physics motivation for a linear collider has been strengthened even more after
the discovery of a new particle compatible with a Higgs boson at the LHC whose
interactions could be studied at the LC with sufficient precision 
to test the Electroweak Symmetry Breaking (EWSB) mechanism of the Standard Model. 
The LC would also probe the dynamics behind the symmetry breaking mechanism
through high-precision measurements of the properties of the top quark,
the heaviest of the fundamental fermions and thus the 
most strongly coupled to the EWSB sector. 
The flexibility in energy of a LC
would allow the top--antitop threshold behaviour to be mapped out in detail. 
Particularly, a theoretically well-defined top mass
with a total uncertainty below 100~MeV could be extracted by means
of such a threshold scan~\cite{Martinez:2002st,Seidel:2013sqa},
substantially beyond the precision achieved at Tevatron,
$m_t=173.18\pm0.56({\rm stat})\pm0.75({\rm syst})$~GeV~\cite{Aaltonen:2012ra},
and at the LHC at 7 TeV, 
$m_t=173.3\pm0.5({\rm stat})\pm 1.3({\rm syst})$~GeV~\cite{ATLAS:2012coa}.
While the statistical uncertainty of the top mass measurement at the
LHC is expected to improve in future runs, the systematic (theoretical)
uncertainties related to the connection between the mass parameter
used in the theory and the one measured in the experiment are
a limiting factor for further improving the accuracy of the top-quark
mass measurement at hadron colliders. 

The perturbative nature of the $t\bar{t}$ system 
which is produced near threshold in $e^+e^-$ annihilation was
recognized long ago~\cite{Bigi:1986jk}, and the leading-order
Coulomb force was treated to all orders in $\alpha_s$
using a non-relativistic approach~\cite{Fadin:1987wz,Fadin:1988fn,Strassler:1990nw}.
The matching between QCD and non-relativistic QCD (NRQCD)~\cite{Caswell:1985ui} provided the necessary pieces
to compute the next-to-next-to-leading order (NNLO) QCD corrections to the 
$t\bar{t}$ production cross section~\cite{Hoang:2000yr} in the region defined by relative
velocities of the top and antitop $v\sim\alpha_s$. In this 
fixed-order approach which achieves a systematic summation of terms 
$\alpha_s^n v^{m+1}$ with $n+m\le k$ at order N${}^k$LO, up to 
N${}^3$LO corrections to resonant $t\bar{t}$ production are 
known~\cite{Beneke:2005hg,Beneke:2008ec,Beneke:2008cr}.
In parallel, the advances in the formulation of the non-relativistic
effective theory allowed for renormalization-group improved calculations for the $t\bar{t}$ system
produced at threshold.
Within the potential NRQCD (pNRQCD)~\cite{Pineda:1997bj,Beneke:1998jj,Beneke:1999qg,Brambilla:1999xf}
and velocity NRQCD (vNRQCD)~\cite{Luke:1999kz,Manohar:1999xd,Hoang:2002yy}
formalisms the systematic summation of potentially large logarithmic
terms $(\alpha_s \log v)^n$
originating from ratios of the top-mass scale $m_t$, the non-relativistic
three-momentum $\vec{p} \sim m_t v$ and the kinetic energy $E\sim m_t v^2$, 
was carried out to next-to-next-to-leading logarithmic (NNLL) 
order~\cite{Hoang:2000ib,Hoang:2001mm,Pineda:2006ri,Hoang:2011it} for the total
cross section, which accounts for all terms proportional 
to $\alpha_s^n v^{m+1}\log^\ell v$ with $n+m-\ell\le 2$.

The effective field theory (EFT) computations above account for the QCD interactions among nearly on-shell
top and antitop quarks. However, the predictions can only be evaluated for all
threshold energies after the top decay width is included in the 
EFT quark propagator, $(E-\vec{p}^2/2m_t +i \Gamma_t)^{-1}$, thus 
providing an infrared cutoff for the top kinetic energy. The counting
$\Gamma_t\sim m_t v^2 \sim m_t \alpha_s^2$ is naturally enforced in this way, 
which is also justified numerically in the Standard Model, where $\Gamma_t \approx 1.5$~GeV
due to the electroweak interaction. 
Once the top width is included, the physical 
final state is $W^+ W^- b\bar b$ -- at least 
if we assume that $V_{tb}\approx 1$, and consider $W$ bosons as 
stable. Beyond leading order, the production of
the final state $W^+ W^- b\bar b$ can also occur through 
non-resonant processes that do not involve a nearly on-shell
$t\bar{t}$ pair, and which are thus not described by the standard
NRQCD formalism. In the counting scheme where the electroweak coupling scales as
$\alpha_{\rm EW}\sim \alpha_s^2$, the leading 
non-resonant effects are NLO for 
the total cross section and reproduce the full-theory contributions
where one of the $bW$ pairs is produced from a nearly on-shell 
top, while the other is produced either from a highly virtual
top or directly without an intermediate top. 
The unstable-particle EFT~\cite{Beneke:2003xh,Beneke:2004km,Beneke:2007zg}
provides the framework for a systematic computation of resonant and non-resonant
contributions while maintaining an expansion in the
small parameters of the problem.
The NLO non-resonant corrections, calculated within this formalism
in~\cite{Beneke:2010mp}, represent the leading electroweak correction
to the $t\bar{t}$ cross section below the threshold, where the LO (resonant) result rapidly vanishes,
reaching up to 20\%. This had been noticed before in Refs.~\cite{Hoang:2008ud,Hoang:2010gu}, obtaining
the dominant NLO non-resonant corrections when moderate invariant-mass cuts
on the $bW$ pairs are applied within the so-called {\it phase-space matching} approach,
based on vNRQCD.

Aside from the sizeable corrections induced by the non-resonant production,
there is a further conceptual reason to term the pure QCD resonant result alone
that is usually shown in the literature as incomplete.
The resonant cross section at NNLO shows {\it finite-width divergences},
{\it i.e.} uncanceled divergences 
proportional to the top width, which in dimensional regularization have the form
\begin{equation}
\label{eq:sigmadiv}
\delta\sigma_{\rm res}^{\mathrm{NNLO}} \propto \frac{\alpha_s \Gamma_t}{\epsilon} \propto \frac{\alpha_s \alpha_{\mbox{\tiny EW}} }{\epsilon}\,,
\end{equation} 
and arise from the logarithmic divergences in the imaginary part of the 
two-loop non-relativistic correlation function. These are 
also known as {\it phase-space divergences} because they can be traced back 
to UV-divergences in the
NRQCD $t\bar{t}$ phase space integrations~\cite{Hoang:2004tg} that 
originate because the unstable-particle propagators 
describing the top quark in the EFT allow for contributions
to the forward-scattering amplitude from 
intermediate top and antitop states which have arbitrarily large invariant masses 
(see \cite{RuizFemenia:2012ma} for a detailed explanation).
The occurrence of finite-width divergences is an evidence that
the pure resonant result must be supplemented with additional short-distance
information from the full $e^+e^- \to W^+W^- b\bar{b}$ process. 
In the unstable-particle
EFT the additional input is given by diagrams corresponding to off-shell top quark
decay that contribute to the non-resonant part at NNLO~\cite{Beneke:2008cr}.
In this paper we show that the NNLO non-resonant contributions 
generate infrared divergences when the momentum of the virtual top-quark lines 
approaches the endpoint at $p_t^2= m_t^2$, which precisely cancel 
the finite-width divergences (\ref{eq:sigmadiv}) from the resonant part.
This is proved by
explicit computation of the NNLO non-resonant contributions,
given by the ${\cal O}(\alpha_s)$ corrections to the NLO non-resonant diagrams.
For the extraction
of the endpoint-singular terms we use the method of regions to asymptotically
expand the loop and phase-space integrals around the endpoint. As a byproduct,
we obtain the first terms in the expansion in $\Lambda/m_t$ of the
NNLO non-resonant contributions, where $\Lambda$ 
is a cut on the invariant masses of the top and antitop decay products
satisfying $m_t \Gamma_t \ll \Lambda^2 \ll m_t^2$. This approximation provides an accurate
estimate of the NNLO non-resonant contributions to the $t\bar{t}$ inclusive cross section
with moderate invariant-mass cuts in the $bW$ systems, and, as we prove,
confirms the result obtained for the same observable within the phase-space matching
approach in~\cite{Hoang:2010gu}.

The structure of the paper is the following. In Section~\ref{sec:EFTdivs} we 
recall the issue of the uncanceled divergences proportional to the top 
width in resonant $t\bar{t}$ production
at NNLO and collect the total divergent result. The framework to 
account for non-resonant production at NLO and NNLO 
is described in Section~\ref{sec:nonresonantdivs}. The origin of the endpoint singularities
that arise in the non-resonant amplitudes and the method used for their analytic extraction 
is explained in the latter Section, which includes the introduction of the scale $\Lambda$.
The results for the endpoint-singular contributions
to the NNLO non-resonant corrections
are summarized in Section~\ref{sec:results} for the various sets of diagrams. 
Readers not interested in the individual results might jump directly to Section~\ref{sec:NNLOres},
where the formula containing all endpoint-singular
non-resonant NNLO contributions to the $e^+e^-\to  W^+W^-b\bar{b}$
cross section with an invariant-mass cut $\Lambda$ is given and the cancellation
of the finite-width divergences against the endpoint divergences is made explicit.
In Section~\ref{sec:comparison} we compare our findings with those of the
phase-space matching approach~\cite{Hoang:2010gu} and also comment on the
approximation to the NNLO non-resonant contributions obtained in another
work~\cite{Penin:2011gg} by expanding in $\rho = 1-M_W/m_t$.
Finally, in Section~\ref{sec:finalresults}
we compare the non-resonant contributions computed in this work with the NLO ones 
as a function of the cut $\Lambda$, and then show their numerical impact 
as a function of the energy relative to the leading-order QCD calculation
of resonant $t\bar{t}$ production. Our conclusions are given in Section~\ref{sec:conclusion}.

\section{NNLO resonant contributions: Finite-width divergences}
\label{sec:EFTdivs}

Close to the top--antitop production threshold, the $W^+W^- b\bar{b}$ final state is produced from $e^+e^-$
collisions predominantly
by intermediate top and antitop quarks with small virtuality (resonant), {\it i.e.} $p_t^2- m_t^2\sim m_t^2 v^2$, where
$v^2=(E+i\Gamma_t)/m_t$ and $E=\sqrt{s}-2m_t$ is the non-relativistic kinetic energy. 
The QCD dynamics of the nearly
on-shell top and antitop quarks can be described within the NRQCD approach, an effective field theory
that is built upon integrating out the hard modes with scale $\sim m_t$. Since the inclusive cross section
for the  $e^+e^- \to W^+W^- b\bar{b}$ process can be obtained from the $W^+W^-b\bar{b}$ cuts of the 
$e^+e^-$ forward-scattering amplitude, the resonant contribution to this observable
is given in the EFT formalism 
by the imaginary part of the matrix element 
\begin{equation}
\label{eq:Ares}
i {\cal A}_{\rm res} = \sum_{k,l} C^{(k)}_p  C^{(l)}_p \int d^4 x \,
\braket{e^- e^+ |
\,\Tprod{\,i {\cal O}_p^{(k)\dagger}(0)\,i{\cal O}_p^{(l)}(x)}\,|e^- e^+} \,,
\end{equation}
where ${\cal O}_p^{(l)}(x)$ (${\cal O}_p^{(k)\dagger}(x)$) are operators
describing the production (decay) of the resonant $t\bar t$ pair from $e^+e^-$, and
$C^{(k,l)}_p$ are short-distance coefficients. At leading order
in the non-relativistic expansion, the $t\bar t$ pair is produced in an $S$-wave, and the 
first term in (\ref{eq:Ares}) is of order $\alpha_{\rm EW}^2 v$. The
corresponding LO production operators can be found in~\cite{Beneke:2010mp}. At NNLO, 
$P$-wave production operators, as well as new ($v^2$-suppressed)
$S$-wave operators contribute (for explicit expressions see \cite{Hoang:2010gu}).
The perturbative contributions to the resonant amplitude~(\ref{eq:Ares}) are
characterized by top and antitop quark lines with time and spatial components
of the momenta 
obeying the potential scaling, $p_t^0-m_t\sim m_t v^2$ and $\vec{p}_t\sim m_t v$
(in the centre-of-mass system or in a reference frame differing from it only
by a small non-relativistic velocity).

\begin{figure}[t]
\begin{center}
\includegraphics[scale=0.75]{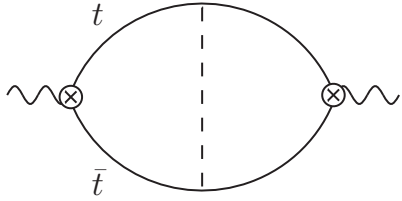}
\caption{Two-loop resonant NRQCD graph with a Coulomb potential.}
\label{figCoulomb}
\end{center}
\end{figure}

As mentioned in the introduction, at NNLO the resonant amplitude~(\ref{eq:Ares}) shows 
an uncanceled finite-width divergence in the imaginary part.
The finite-width  divergences arise from two-loop 
diagrams with a Coulomb gluon and a $v^2$-suppressed insertion. The insertion may
correspond to a $v^2$-suppressed potential, an NNLO correction to the kinetic Lagrangian,
a $P$-wave or a $v^2$-suppressed $S$-wave current producing the top--antitop pair, 
or an insertion of the absorptive part of matching coefficients of production operators
describing finite lifetime corrections. The different contributions to the NNLO finite-width 
divergence  are collected in the following formula:

\begin{align}
{\rm div} \, \sigma^{\rm NNLO}_{\rm res} = & \,\,
  \left[ \big( C_{p}^{(v)} \big)^2+ \big( C_{p}^{(a)} \big)^2\right]  2\,N_c\,
  \Big( \, 
    -4\pi a \, {\rm div}\big[ {\rm Im}\, G_r \big]
    + {\rm div}\big[{\rm Im}\, G_{\rm kin} \big]
\nonumber \\
&
    \hspace{4.9cm} + {\rm div}\big[{\rm Im}\, G_{\rm dil}\big] + {\rm div}\big[{\rm Im}\, G_{v^2}\big]
  \Big)
\nonumber \\
&
  + \left[\big(C_{p,P\rm \text{-wave}}^{(v)} \big)^2 + \big( C_{p,P\rm \text{-wave}}^{(a)} \big)^2\right] 
   \frac{4N_c}{3 m_t^2} \,{\rm div}\big[{\rm Im}\,G_{P\text{-wave}} \big]
\nonumber \\
&
  + \left[ C_p^{(v)}C_p^{(v),\rm abs}+ C_p^{(a)}C_p^{(a),\rm abs}\right] 4 N_c
  \,{\rm div}\big[{\rm Re}\,G^{(0)}_{\rm C} \big] \, ,
\label{divsigma}
\end{align}
where
\begin{eqnarray}
C_p^{(v)} = 4\pi\alpha
\,\left[ \frac{Q_t Q_e}{s} + \frac{v_e v_t}{s-M_Z^2} \,
\right] \quad , \quad
C_p^{(a)} = -4\pi\alpha
\,\frac{a_e v_t}{s-M_Z^2} \,,
\\
 C_{p,P\rm \text{-wave}}^{(v)} = 4\pi\alpha
\, \frac{v_e a_t}{s-M_Z^2} \quad , \quad
C_{p,P\rm \text{-wave}}^{(a)} = -4\pi\alpha
\,\frac{a_e a_t}{s-M_Z^2}
\end{eqnarray}
are the tree-level coefficient functions of the leading-order $S$- and $P$-wave
production operators, and $a\equiv \CF\,\alpha_s$, with
$\CF=4/3$ the Casimir operator of the fundamental  
SU(3) representation.
The vector ($v_f$) and axial-vector ($a_f$) couplings of the fermions
to the gauge bosons are given in Eq.~(\ref{eq:vfaf}) below, 
and the remaining symbols are defined as in~\cite{Beneke:2010mp}.
The $C_p^{(v/a),\rm abs}$ are the absorptive parts of the matching coefficients of
the leading-order $S$-wave production operators. The latter contributions are part of the 
corrections induced by the top-quark instability, as they can be interpreted
as top-pair production where one of the tops arises from a $bW$ system with invariant mass very close
to $m_t^2$. The coefficients $C_p^{(v),\rm abs}$ and $C_p^{(a),\rm abs}$
were calculated in Ref.~\cite{Hoang:2004tg} (there denoted 
$C_V^{bW,\rm abs}$ and $C_A^{bW,\rm abs}$, respectively), and their
explicit expressions can be found therein. They are suppressed by $\alpha_{\rm EW}\sim \Gamma_t/m_t$
with respect to the leading-order $C_p^{(v/a)}$,
and therefore amount to NNLO corrections. The absorptive parts in the matching coefficients
of the $t\bar{t}$ production currents lead to a dependence of the
cross section on the real part of the zero-distance
Coulomb Green function $G^{(0)}_{\rm C}$, which has a divergence
from the two-loop graph with a Coulomb potential shown in Fig.~\ref{figCoulomb},
\begin{eqnarray}
{\rm div}\big[{\rm Re}\,G^{(0)}_{\rm C} \big] = \frac{m^2_t a}{16\pi}\,\frac{1}{\epsilon} \,,
\label{divReGc}
\end{eqnarray}
here regulated dimensionally. 

The remaining contributions to ${\rm div} \, \sigma^{\rm NNLO}_{\rm res}$ in
(\ref{divsigma}) 
arise from divergent contributions to the imaginary part of the Green function:
$G_r$, $G_{\rm kin}$ and $G_{\rm dil}$ are first-order corrections to the 
zero-distance Green function from the potential
$\widetilde{V}_r(\vec{p},\vec{q})=(\vec{p}^2+\vec{q}^2)/2m_t^2(\vec{p}-\vec{q})^2$
(in momentum space, leaving out the corresponding Wilson coefficient), 
the insertion
of the kinetic energy correction $\vec{\partial}^4/(8m_t^3)$, 
and the insertion of the lifetime-dilatation
operator 
$i\Gamma_t\vec{\partial}^{\,2}/(4m_t^2)$, respectively. The explicit expressions 
for $G_r$, $G_{\rm kin}$ and $G_{\rm dil}$ in $d=4-2\epsilon$ dimensions 
can be found in Ref.~\cite{Hoang:2004tg}.
Their divergent parts satisfy
\begin{eqnarray}
{\rm div}\big[{\rm Im}\,G_{\rm kin} \big] = -{\rm div}\big[{\rm Im}\,G_{\rm dil} \big]
=-(4\pi a) \,{\rm div}\big[{\rm Im}\,G_{r} \big] = 
\frac{m_t a}{16\pi}\,\frac{\Gamma_t}{\epsilon} \,.
\label{divReGrGkinGdil}
\end{eqnarray}
The term $G_{v^2}$ corresponds to the
zero-distance Coulomb Green function obtained with the top--antitop
pair produced by the $v^2$-suppressed $S$-wave current, which can easily be
related to the leading-order Coulomb Green function by the 
non-relativistic equation of motion of the top quark, $G_{v^2} = -(v^2/3) G^{(0)}_{\rm C}$. Therefore,
\begin{eqnarray}
{\rm div}\big[{\rm Im}\,G_{v^2} \big] = -\frac{\Gamma_t}{3m_t} {\rm div}\big[{\rm Re}\,G^{(0)}_{\rm C} \big]
= -\frac{m_t a}{48\pi}\,\frac{\Gamma_t}{\epsilon} \,.
\label{divImGv2}
\end{eqnarray}
Finally, $G_{P\text{-wave}}$ is the $\ell=1$ component of the Coulomb Green
function at zero distance and describes the production of the top--antitop pair
in a $P$-wave, which first contributes at NNLO. From the explicit expression of
$G_{P\text{-wave}}$ (see {\it e.g.} Ref.~\cite{Hoang:2004tg}, denoted $G^1$ therein)
we have
\begin{eqnarray}
{\rm div}\big[{\rm Im}\,G_{P\text{-wave}} \big] 
= \frac{m_t^3 a}{16\pi}\,\frac{\Gamma_t}{\epsilon} \,.
\label{divImGPwave}
\end{eqnarray}
Using the results (\ref{divReGc})--(\ref{divImGPwave}) in (\ref{divsigma}),
we get the total finite-width divergence at NNLO: 
\begin{align}
{\rm div} \, \sigma^{\rm NNLO}_{\rm res} = & \,\,
   \frac{ C_F N_c \, \alpha_s}{12\pi} \frac{m_t\Gamma_t}{\epsilon} \, \Big[
     \big( C_{p}^{(v)} \big)^2+ \big( C_{p}^{(a)} \big)^2
   + \big(C_{p,P\rm \text{-wave}}^{(v)} \big)^2 + \big( C_{p,P\rm \text{-wave}}^{(a)} \big)^2
\nonumber \\*
   & \hspace{3.5cm} + \frac{3m_t}{\Gamma_t}\left( C_p^{(v)}C_p^{(v),\rm abs}+ C_p^{(a)}C_p^{(a),\rm abs}\right)
  \Big]\,.
\label{divsigmatotal}
\end{align}
The result in (\ref{divsigmatotal}) agrees with the total UV divergence generated in the NNLO effective-theory
matrix elements contributing to $\sigma^{\rm NNLO}_{\rm res}$ as obtained in~\cite{Hoang:2004tg}. 
We note that in the approach followed in Refs.~\cite{Hoang:2010gu,Hoang:2004tg}, the finite-width divergences (named
phase-space divergences therein) are absorbed by the counterterms $\delta\tilde{C}_{V/A}$ associated to $(e^+e^-)(e^+e^-)$
for\-ward-scattering operators, and the corresponding phase-space logarithms are resummed using renormalization-group 
techniques known from effective theories for the coefficients $\tilde{C}_{V/A}$. The matching conditions for these
coefficients at the hard scale, $\tilde{C}_{V/A}(\nu=1)$, are related to the non-resonant contributions in the unstable-particle
effective theory approach which we discuss in the following section.

\section{Endpoint divergences in the non-resonant contributions}
\label{sec:nonresonantdivs}

The non-relativistic EFT formalism has to be extended in order to account for non-resonant production of the 
physical final state $W^+W^- b\bar{b}$, which involves processes where the $bW$ pairs are produced 
by highly virtual top or antitop quarks, $p_t^2- m_t^2\sim {\cal O}(m_t^2)$, or without intermediate tops. 
The leading non-resonant contributions are caused by the top-quark instability, and are thus
of electroweak origin. Adopting the counting scheme where $\alpha_{\rm EW}\sim \alpha_s^2$,
a systematic separation of resonant and non-resonant
effects can be achieved within the unstable-particle effective theory for pair production 
near threshold~\cite{Beneke:2003xh,Beneke:2004km,Beneke:2007zg,Hoang:2004tg}. 
Non-resonant effects take place at short distances, as compared to the 
length scales governing the QCD interaction of the non-relativistic top--antitop pair. The 
non-resonant contributions to the $e^+e^-$ forward-scattering amplitude are thus reproduced
in this formalism by the matrix element of four-electron production--decay operators 
${\cal O}_{4e}^{(k)}(0)$~\cite{Beneke:2003xh,Beneke:2004km,Hoang:2004tg}:
\begin{equation}
\label{eq:Anonres}
i {\cal A}_{\rm non-res} = \sum_{k} \,C_{4 e}^{(k)} 
\braket{e^- e^+|i {\cal O}_{4e}^{(k)}(0)|e^- e^+}\,,
\end{equation}
with short-distance coefficients  $C_{4 e}^{(k)}$ that are determined by the hard contributions
of the $e^+e^-$ forward-scattering amplitude. Explicit expressions for the operators
${\cal O}_{4e}^{(k)}$ can be found in~\cite{Beneke:2010mp}. 

\begin{figure}[t]
\begin{center}
\includegraphics[width=0.9\textwidth]{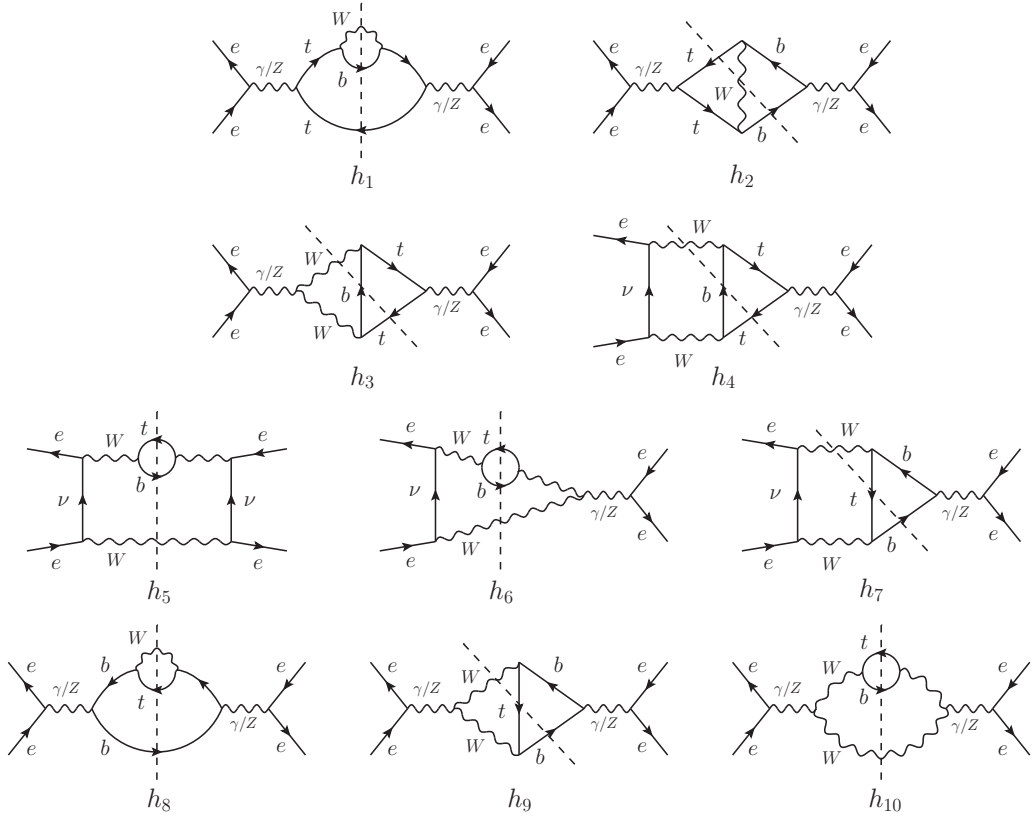}
\caption{Two-loop forward-scattering amplitude diagrams with 
$b W^+ \bar{t}$ cuts contributing to the NLO non-resonant cross section. $\bar{b} W^- t$ cuts and symmetric diagrams 
are not shown. This figure is reproduced from Ref.~\cite{Beneke:2010mp}.}
\label{fig1}
\end{center}
\end{figure}

For the $W^+W^- b\bar{b}$ inclusive cross section,
only the imaginary parts of $C_{4 e}^{(k)}$ are needed. The first non-vanishing contribution 
arises from the cut two-loop diagrams of order $\alpha_{\rm EW}^3$ shown in Fig.~\ref{fig1}, and is
thus suppressed by $\alpha_{\rm EW}/v\sim v$ (NLO) with respect to the leading term from 
the resonant part.\footnote{We do not consider other ${\cal O}(\alpha_{\rm EW}^3)$ contributions
where the $b\bar{b}$ or the $W^+W^-$ pair in the final state is reached through the resonant decay of 
a Higgs or $Z$-boson, since the latter constitute a reducible background which
can be eliminated in the $t\bar{t}$ resonance region by applying cuts on the invariant masses of the 
final-state particle pairs~\cite{Beneke:2010mp}.}
The diagrams in Fig.~\ref{fig1} are obtained by asymptotically expanding the full electroweak theory diagrams 
of the $e^+e^-$ forward-scattering amplitude 
assuming that the momenta in the top and antitop lines are hard, so that the top quarks are far off-shell,
$p_t^2-m_t^2\sim {\cal O}(m_t^2)$. This expansion implies that the top-quark self-energy insertions
(resummed in the full-theory diagrams in the top-quark propagator),
$m_t \Sigma(p_t^2)\sim m_t^2 \alpha_{\rm EW} \ll p_t^2-m_t^2$, must be treated perturbatively. Accordingly, 
top-quark propagators in the non-resonant diagrams have  no width, contrary to the case of 
resonant top quarks. 
The leading-order imaginary contribution in this non-resonant expansion of
the top propagator is proportional to $\delta(p_t^2-m_t^2)$, yielding the
3-particle final states $b W^+ \bar{t}$ (Fig.~\ref{fig1}) and
$\bar{b} W^- t$ instead of the physical 4-particle final state
$W^+W^- b\bar{b}$.
In addition, the 
amplitudes corresponding to the diagrams in Fig.~\ref{fig1} have to be expanded in $v$
in the threshold region.
Altogether, the non-resonant contribution at NLO
amounts to the calculation of the squared and phase-space integrated amplitudes for 
the on-shell processes $e^+e^- \to \bar{b} W^- t$ and $e^+e^- \to b W^+ \bar{t}$
at the centre-of-mass energy $s=4m_t^2$ in ordinary perturbation theory
(see~\cite{Beneke:2010mp}).
The NLO non-resonant corrections to the  $W^+W^- b\bar{b}$ cross section, also including 
cuts on the invariant masses of the $bW$ systems, were determined in~\cite{Beneke:2010mp} and
later confirmed by~\cite{Penin:2011gg}.

Divergences in the non-resonant part can arise 
when the top (or antitop) propagators go 
on-shell~\cite{Beneke:2010mp,RuizFemenia:2012ma}. This is a consequence of the hard-momentum
region expansion, which forces us to drop the top width from the top--antitop propagators.
Given that dimensional regularization is used to deal with divergences in the 
resonant amplitude, it must also be used here consistently in order to regulate phase-space singularities 
from top-quark propagators going on-shell. 
For the 3-particle $b W^+ \bar{t}$ phase space, the outer integration variable can be chosen 
as the squared invariant mass of the $b W^+$ subsystem, $p_t^2$, where $p_t = p_b + p_W$. The antitop momentum is on-shell here, $p_{\bar t}^2 = m_t^2$.
Setting $s=4m_t^2$, as dictated by the asymptotic expansion, the kinematics 
of the process provides the restriction $m_t^2-\Lambda^2\le p_t^2 \le m_t^2$, where
$\Lambda^2=2m_t \Delta M_t-\Delta M_t^2$ is introduced to allow for loose cuts
($\Lambda^2 \gg m_t \Gamma_t$ or $\Delta M_t \gg \Gamma_t$)
in the $b W$ invariant masses of the form~\cite{Beneke:2010mp}
\begin{align}
m_t -\Delta M_t &  \le \sqrt{p_{t,\bar{t}}^2} \le m_t +\Delta M_t\,.
\label{eq:DeltaM}
\end{align}
For the $\bar{b} W^- t$ phase space, the roles of $p_t^2$ and $p_{\bar t}^2$ are
reversed.

To recover the total cross section we have to set $\Lambda^2=m_t^2-M_W^2$. 
Integrating over all other kinematic variables but $p_t^2$, each non-resonant contribution
involves an integral of the form 
\begin{equation}
\int^{m_t^2}_{m_t^2-\Lambda^2} \frac{dp_t^2}{(m_t^2-p_t^2)^{r+ n \epsilon}} \,
 = \, \frac{1}{1-r-n\epsilon}\, (\Lambda^2)^{1-r-n\epsilon} \,,
\label{eq:endpointsing}
\end{equation}
where the endpoint singularity at $p_t^2=m_t^2$ for $r\ge 1$  has been regularized in $d=4-2\epsilon$ dimensions,
which drops the scaleless singular contribution from the upper boundary $p_t^2=m_t^2$.
The integrals~(\ref{eq:endpointsing}) arise from expanding the non-singular
parts of the numerator of the contributions about the endpoint $p_t^2=m_t^2$.
At NLO, only the diagram $h_1$ has an endpoint divergence, with $r=3/2$, and
the dimensionally regularized result is therefore finite 
in the limit $\epsilon \to 0$. At NNLO, however, integrands with $r=1$ are found which generate $1/\epsilon$ terms 
of the form (\ref{eq:sigmadiv}). These shall cancel the finite-width divergences in the resonant part of the full-theory
diagrams.

{}From the unstable-particle EFT power-counting $\alpha_s\sim \alpha_{\rm EW}^{1/2}$, NNLO non-resonant
corrections can only arise from QCD corrections to the NLO ones. More precisely,
the non-resonant NNLO corrections  are obtained from the
NLO contributions by adding to the diagrams $h_1$--$h_{10}$ in Fig.~\ref{fig1}
a virtual gluon or QCD counterterm (on one side of the cut) or a real gluon (going through the cut) 
in all possible ways. The number of diagrams contributing in this way is well above 100. Fortunately, only
a few of them are endpoint-singular, as we show next, first for the virtual contributions and afterwards
for the real-gluon radiation diagrams.

\begin{figure}[t]
\begin{center}
\includegraphics[width=0.24\textwidth]{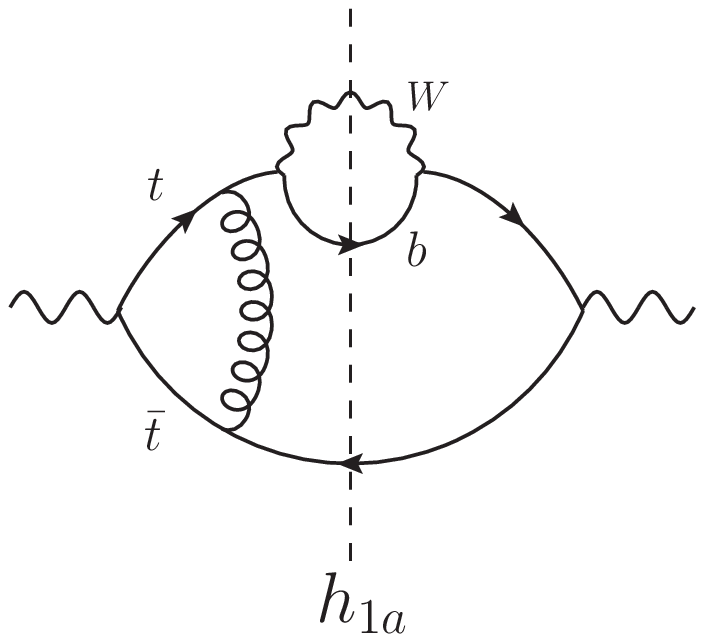}
\hspace*{0.cm}
 \includegraphics[width=0.24\textwidth]{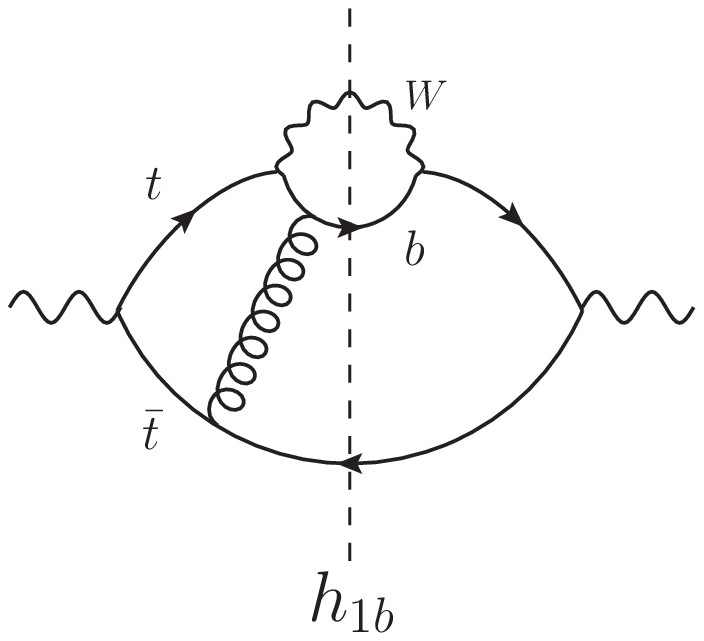}
\hspace*{0.cm}
\includegraphics[width=0.24\textwidth]{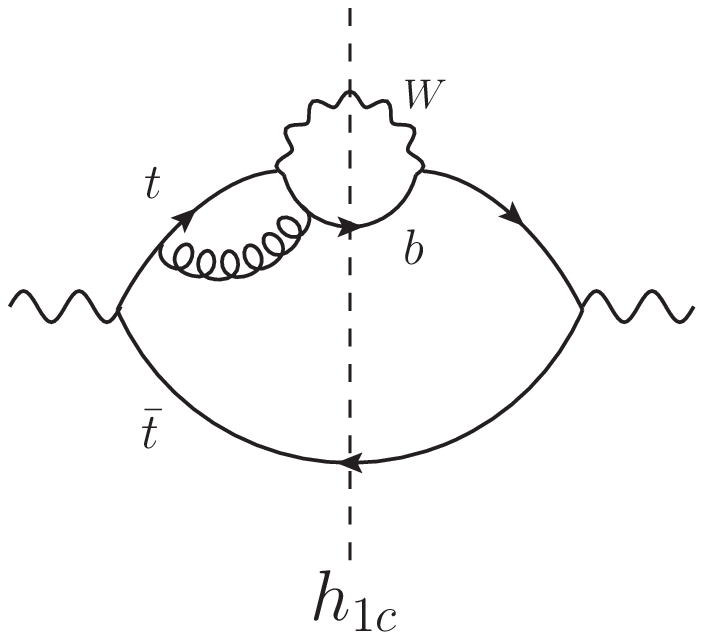}

\vspace*{.2cm}
\includegraphics[width=0.24\textwidth]{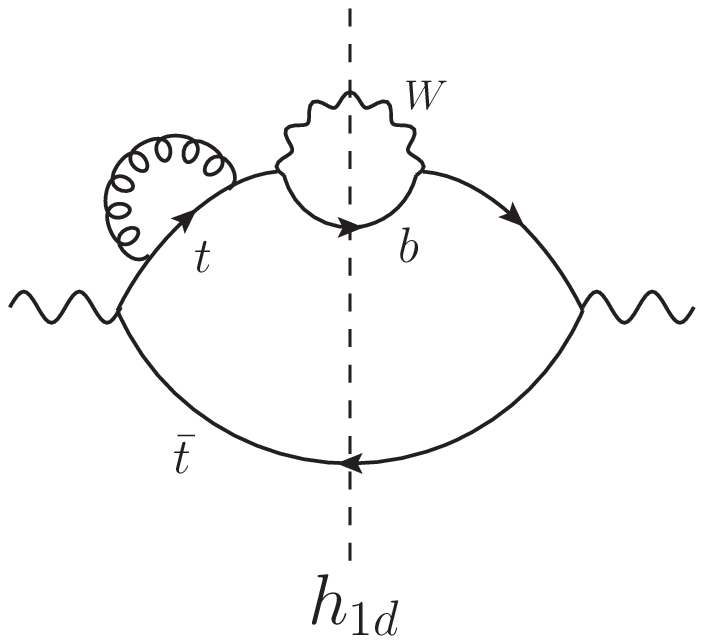}
\hspace*{-.2cm}
 \includegraphics[width=0.24\textwidth]{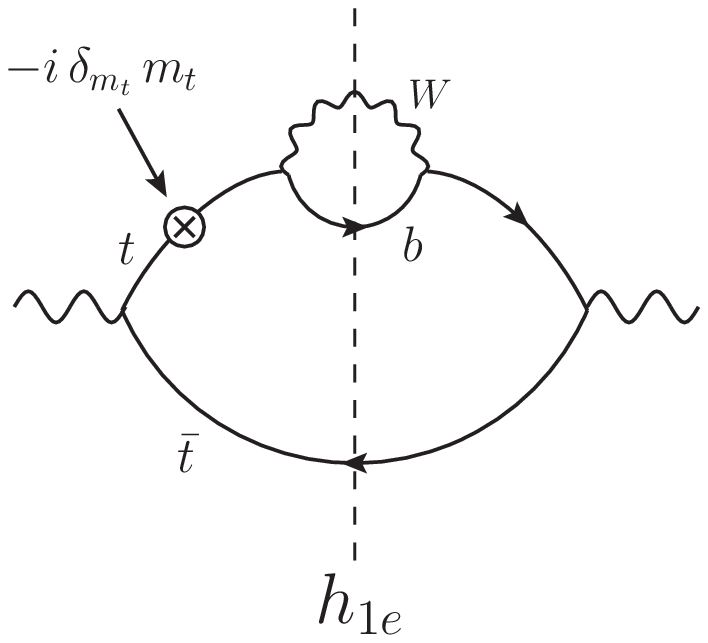}
\hspace*{-.2cm}
\includegraphics[width=0.24\textwidth]{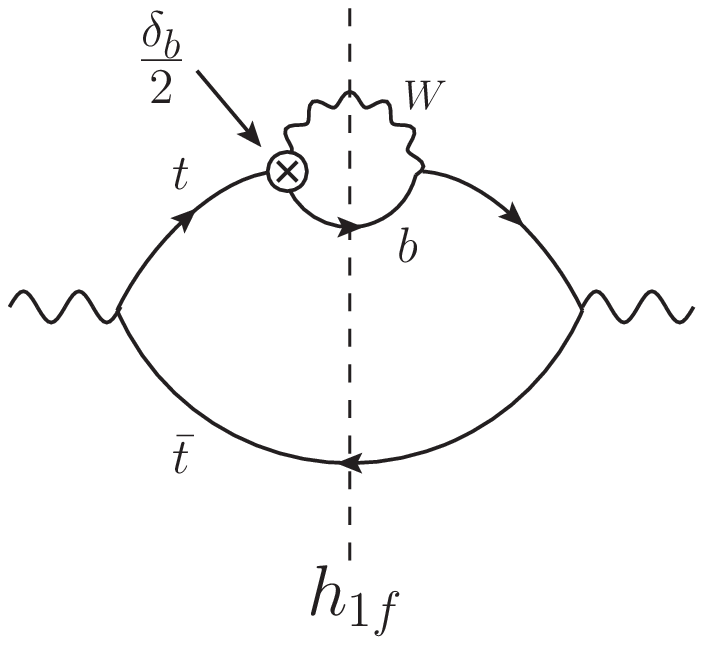}
\hspace*{-.2cm}
\includegraphics[width=0.24\textwidth]{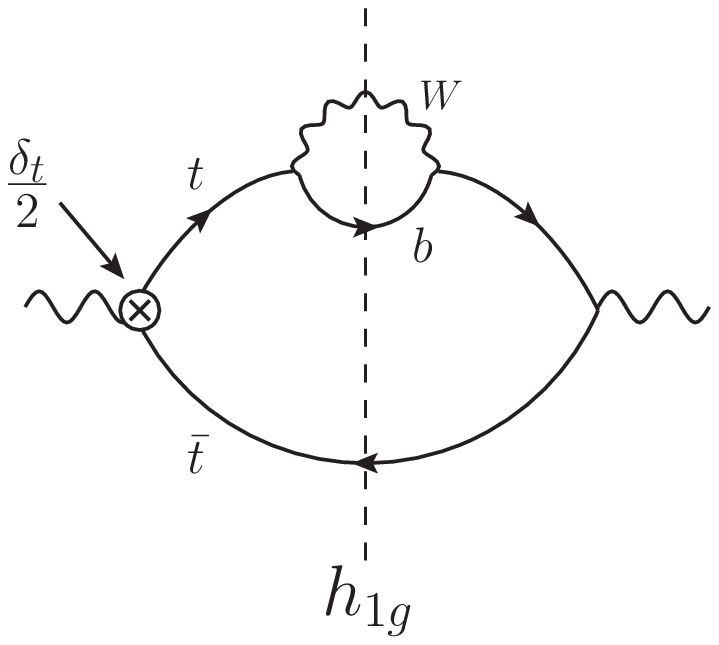}

\vspace*{.2cm}
\includegraphics[width=0.26\textwidth]{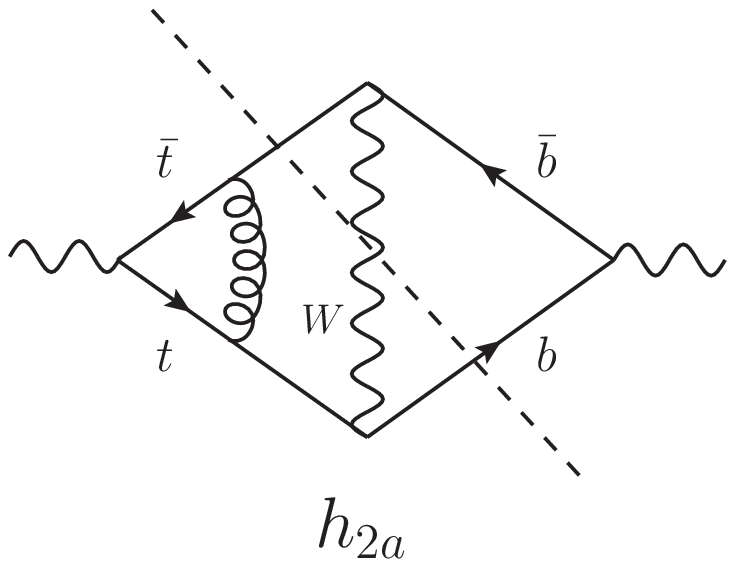}
\hspace*{-.cm}
 \includegraphics[width=0.26\textwidth]{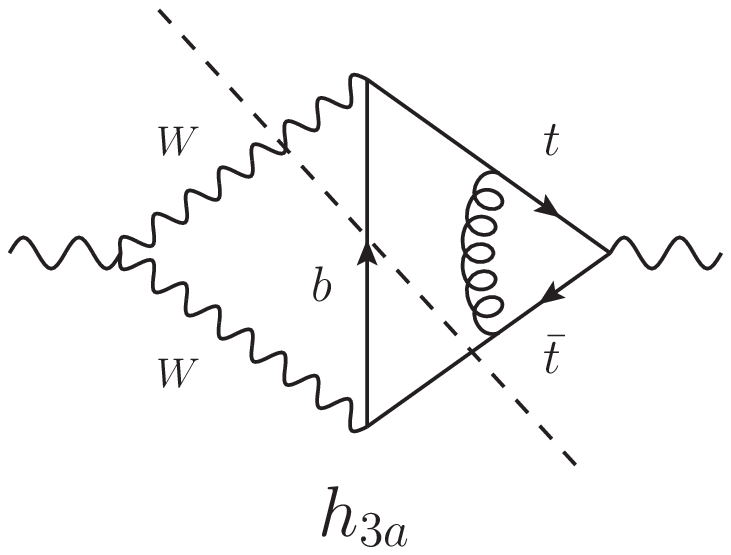}
\hspace*{.0cm}
\includegraphics[width=0.285\textwidth]{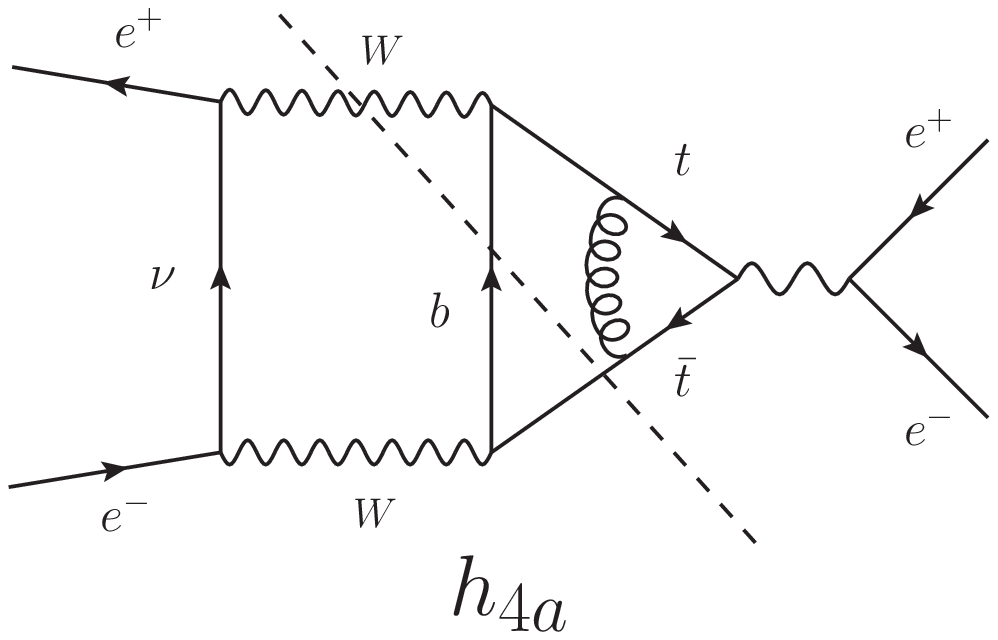}

\caption{${\cal O}(\alpha_s)$ virtual corrections to the two-loop forward-scattering  diagrams with 
$b W^+ \bar{t}$ cuts which are endpoint-singular. $\bar{b} W^- t$ cuts and symmetric diagrams 
are not shown. The $e^+e^-$ external legs have not been drawn, except for the $h_{4a}$ diagram. }
\label{fig2}
\end{center}
\end{figure}

NNLO amplitudes with virtual gluons (see Fig.~\ref{fig2}) involve the same 3-particle 
cuts ($b W^+ \bar{t}$ or $\bar{b} W^- t$) as the NLO diagrams. 
In order to isolate the endpoint-singular contributions, the 
integrand of the $p_t^2$-integration in the 3-particle
phase space is asymptotically expanded in powers of $(1-t)$, where $t=p_t^2/m_t^2$. A
factor $(1-t)^{1/2-\epsilon}$ is provided by the kinematics of the phase space
and originates from splitting the total momentum $q$ into $p_t$ and the
on-shell antitop momentum $p_{\bar t}$. Negative integer powers of $(1-t)$ are
introduced by top propagators. This simple counting provides the leading term
$(1-t)^{-3/2-\epsilon}$ in the limit $t\to 1$ for the integrand of the NLO diagram
$h_1$, while diagrams $h_2\,$--$\,h_4$, with one intermediate top propagator less, 
get an additional power of $(1-t)$ and are endpoint-regular. Likewise, the
integrands for diagrams $h_5\,$--$\,h_{10}$, with no internal top line, exhibit the overall 
scaling $(1-t)^{1/2-\epsilon}$ and are well-behaved at the endpoint.

Gluon insertions can give rise to an additional power $(1-t)^{r^\prime}$, where 
only $r^\prime \le 1/2$ is relevant for an endpoint-singular behaviour, and we note that
half-integer values of $r^\prime$ are needed in order to produce $1/\epsilon$ endpoint
divergences. The overall power in $(1-t)$ of the loop integrals involved in the virtual
corrections can easily be inferred by using the expansion by
regions~\cite{Beneke:1997zp,Smirnov:2002pj,Jantzen:2011nz},
as it is explained in Sec.~\ref{sec:calcmethod} below.
Negative powers of $(1-t)$ can arise from 1-loop 
integrals which (after integration over all phase-space variables but $t$)
are singular when $t$ approaches its maximum. 
This is only achieved by the gluon correction to the $t\bar{t}$-vertex and by the 
virtual gluon connecting the antitop with the on-shell bottom quark 
radiated from a top line. 
The former yields negative powers with
$r^\prime=-1/2-\epsilon$ and thus produces endpoint-singular contributions when 
it corrects diagrams $h_1\,$--$\,h_4$ (NNLO diagrams $h_{ia},\,i=1,\dots,4$ of
Fig.~\ref{fig2}).
In particular,
the negative half-integer power introduced by the $t\bar{t}$-vertex correction 
in the NNLO diagrams $h_{1a}\,$--$\,h_{4a}$ gives integrands scaling as $(1-t)^{-2-2\epsilon}$
(for $h_{1a}$) and $(1-t)^{-1-2\epsilon}$ (for $h_{ia},\,i=1,\dots,4$), where the latter 
produce $1/\eps$ endpoint divergences according to (\ref{eq:endpointsing}).
On the other hand, the virtual gluon connecting the antitop with the on-shell bottom quark 
radiated from a top line
yields factors $(1-t)^{-1-2\epsilon}$ and $(1-t)^{-1/2-\epsilon}$.
However, this loop integral replaces one top propagator of the corresponding
NLO diagram, so the additional powers are $r^\prime=-2\epsilon$ and
$r^\prime=1/2-\epsilon$.
Thus this loop integral is only of relevance when inserted
into diagram $h_1$ (NNLO diagram $h_{1b}$ of Fig.~\ref{fig2}),
yielding singular integrand terms scaling as $(1-t)^{-3/2-3\epsilon}$ and $(1-t)^{-1-2\epsilon}$, respectively.

The remaining virtual $\alpha_s$-corrections to diagram $h_1$, namely the top-bottom vertex corrections (in diagram $h_{1c}$),
the top self-energy plus the mass counterterm ($h_{1d}$ plus $h_{1e}$), as well as the $tb$- and $t\bar{t}$-vertex counterterms ($h_{1f}$ and $h_{1g}$), 
yield factors $(1-t)^{0+n\eps}$ at leading order and $(1-t)^{1+n\eps}$ at
next-to-leading order, such that the corresponding diagrams
have the same degree of divergence at the endpoint as the NLO diagram $h_1$, and subleading terms are endpoint-regular. The same applies to
the remaining virtual corrections to diagrams $h_{i},\,i=2,\dots,10$, which are
not shown in Fig.~\ref{fig2}.
In summary, the potentially endpoint-singular cases that can be found from a virtual gluon at NNLO 
are ($y\equiv1-\Lambda^2/m_t^2$):
\begin{align}
  \label{eq:intep2}
  \int_y^1\! d t \, (1-t)^{-2-2\epsilon}
    &\propto  \, \frac{m_t^2}{\Lambda^2}
      \left(\frac{m_t^2}{\Lambda^2}\right)^{2\epsilon} \,, \quad {\rm diagram}\; h_{1a} \,,
\\
  \label{eq:intep32}
  \int_y^1 \! d t \, (1-t)^{-3/2-n\epsilon}
    &\propto  \, \frac{m_t}{\Lambda}
      \left(\frac{m_t^2}{\Lambda^2}\right)^{n\epsilon} \,, \quad {\rm diagrams}\; h_{1X},\, X=a,\dots,g \,,
\\
  \label{eq:intep1}
  \int_y^1\! d t \, (1-t)^{-1-n\epsilon}
    &\propto  \frac{1}{\epsilon} \left(\frac{m_t^2}{\Lambda^2}\right)^{n\epsilon}
    \simeq \frac{1}{\,\epsilon_\ep} +n \ln\frac{m_t^2}{\Lambda^2} 
    \;\;, \, \; {\rm diagrams} \; h_{ia},\, i=1,\dots,4 ,\;h_{1b}\,.
\end{align}
Only in the last case~(\ref{eq:intep1}) we obtain a $1/\eps$ singularity, which is labeled
$1/\eps_\ep$ in order to mark it as an endpoint divergence.
The cases (\ref{eq:intep2}) and (\ref{eq:intep32}) are also endpoint-divergent,
but finite (in the limit $\epsilon \to 0$) through dimensional regularization.
The asymptotic expansion of the phase-space integral near the endpoint
translates into an expansion in powers of $\Lambda/m_t$ in the results.
Virtual corrections to diagrams $h_{5}\,$--$\,h_{10}$ are all endpoint-regular
and contribute first
at ${\cal O}(\left(\Lambda/m_t\right)^3)$ in this expansion. 

\begin{figure}[t]
\begin{center}
\includegraphics[width=0.24\textwidth]{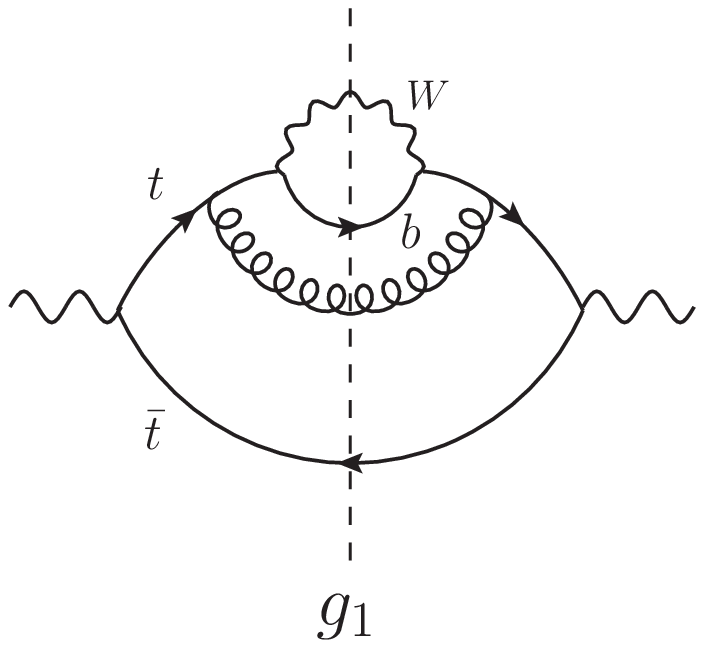}
\hspace*{0.2cm}
 \includegraphics[width=0.24\textwidth]{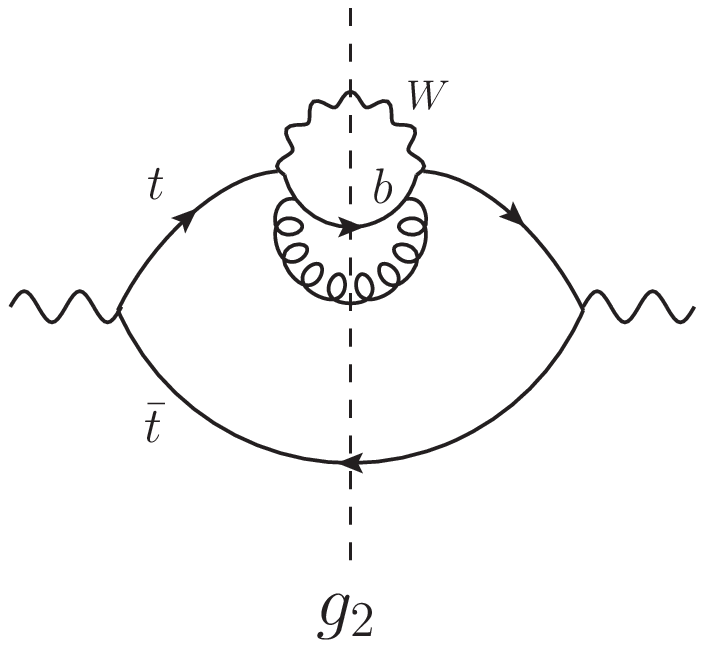}
\hspace*{0.2cm}
\includegraphics[width=0.24\textwidth]{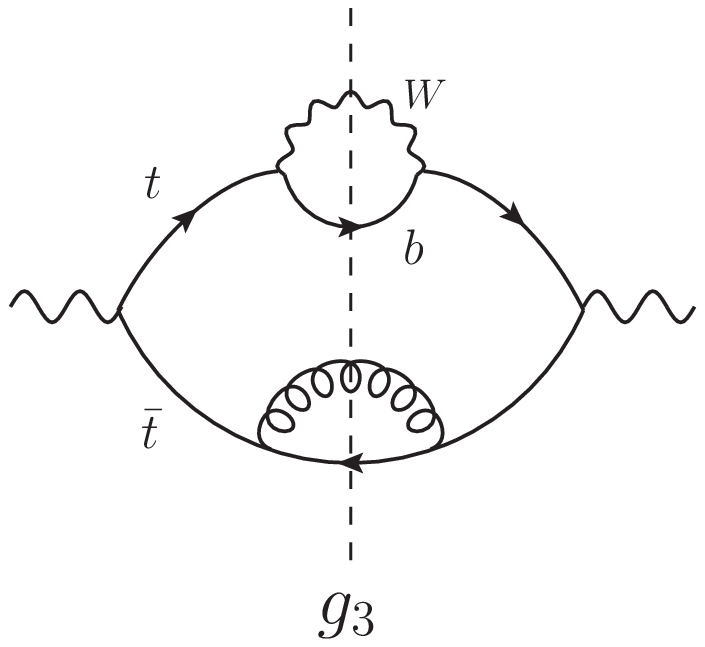}

\vspace*{.2cm}
\includegraphics[width=0.24\textwidth]{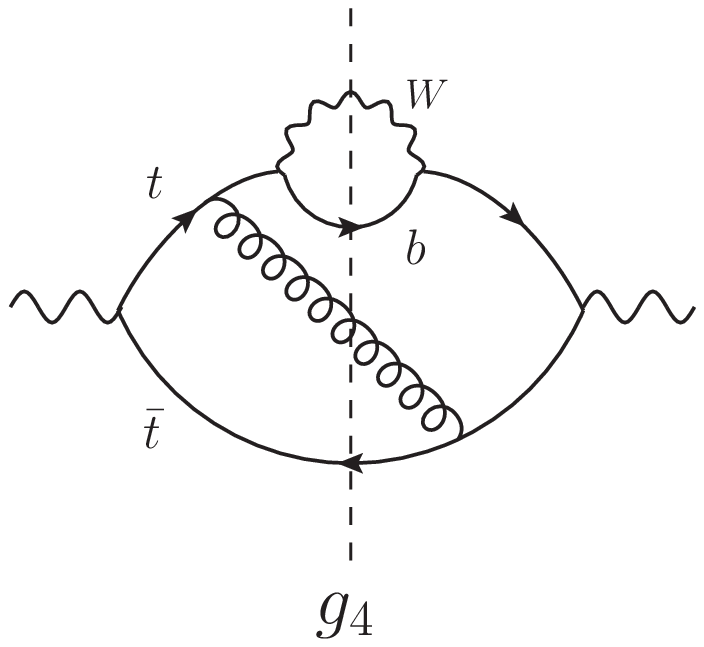}
\hspace*{.2cm}
 \includegraphics[width=0.24\textwidth]{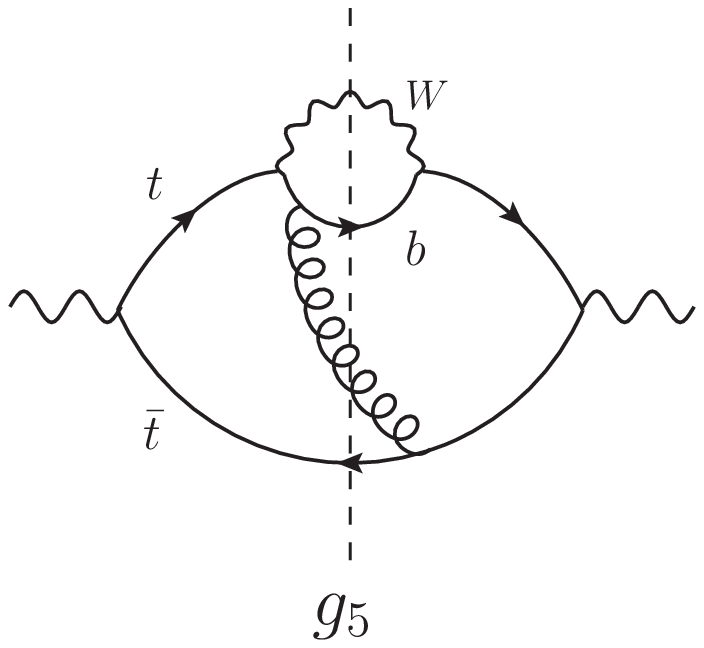}
\hspace*{.2cm}
\includegraphics[width=0.24\textwidth]{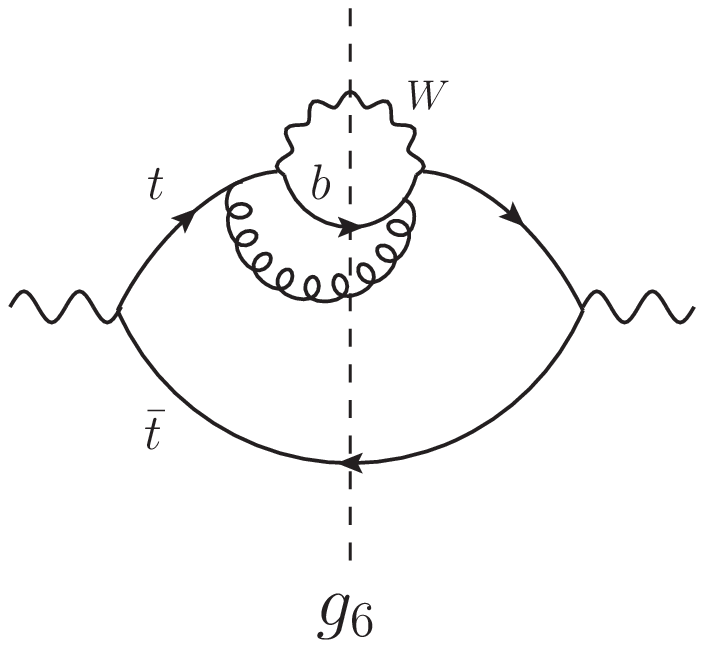}

\caption{${\cal O}(\alpha_s)$ real-gluon corrections to the two-loop forward-scattering  diagrams with 
$b W^+ \bar{t} g$ cuts which are endpoint-singular. $\bar{b} W^- t g$ cuts and symmetric diagrams 
are not shown. The $e^+e^-$ external states have been omitted. }
\label{fig3}
\end{center}
\end{figure}

Let us turn to the NNLO non-resonant contributions with real-gluon emission
which involve a 4-particle cut, $b W^+ \bar t g$ (see Fig.~\ref{fig3}) or
$\bar{b} W^- t g$.
The corresponding phase space can be written with an additional
integration over the variable $t^* = p_{t^*}^2/m_t^2$ with $p_{t^*} = p_b +
p_W + p_g$, where $p_g$ is the gluon momentum. 
The kinematic restriction for the total cross section
reads $x \le t \le t^* \le 1$, where $x=M_W^2/m_t^2$. 
Now we want to impose an invariant-mass cut in accordance with the one for
the 3-particle phase space described above. An infrared-safe definition
requires that the case where no gluon is emitted and the case
where the bottom quark emits a collinear gluon should be indistinguishable.
Note that when the bottom quark emits a collinear
gluon, splitting its momentum into $p_b + p_g$, we may have $t =
(p_b+p_W)^2/m_t^2$ significantly smaller than $t^* = (p_b+p_g+p_W)^2/m_t^2$.
Without real-gluon emission, the invariant-mass cut defined above imposes a
lower bound on the variable~$t$, \textit{i.e.} $t > y$ with $y =
1-\Lambda^2/m_t^2$. With gluon emission from the bottom quark,
the new variable~$t^*$ represents the quantity which before was described
by the variable~$t$. Then, for consistency, the lower cut has to be imposed
on~$t^*$, \textit{i.e.} $t^* > y$.
If, instead, we imposed a lower bound on~$t$ in the real-gluon case, the
situation $t < y < t^*$ could arise, thus excluding events where the gluon
is detected separately ($t < y$) and including events where the gluon is
part of the bottom jet ($t^* > y$).
Therefore, a collinear-safe observable is only obtained 
by imposing the lower limit on the variable~$t^*$
which corresponds to the
3-particle variable~$t$ when the collinear gluon is not detected separately
in the detector.
With this choice, we obtain integrals for the 4-particle phase space
parametrized as 
\begin{align}
 \int_{y}^1 d t^* \int_x^{t^*} d t
 \,.
\end{align}
To extract the endpoint behaviour, these integrals are asymptotically expanded
for small $(1-t^*)$, using the strategy of regions as explained in
Sec.~\ref{sec:calcmethod} below.
In this parameterization, the size of the variable $(t^*-t)$,
which is proportional to the components of~$p_g$, characterizes the
different regions for expanding the inner integral in the limit $t^*\to 1$.
We therefore  need to keep track of
factors of $(1-t^*)$ and $(t^*-t)$ (or of $(1-t)=(1-t^*)+(t^*-t)$)
appearing in the amplitudes. 
The phase-space splitting of the total momentum
into $p_{t^*}$ and the antitop momentum~$p_{\bar t}$ provides a factor
$(1-t^*)^{1/2-\eps}$, and the gluon emission, $p_{t^*} \to p_t + p_g$, yields
$(t^*-t)^{1-2\eps}$. Top propagators with momenta $p_t$ and $p_{t^*}$
contribute negative integer powers of $(1-t)$ and $(1-t^*)$,
respectively, whereas bottom propagators with momentum $(p_b+p_g)$ and
antitop propagators with momentum $(p_{\bar t}+p_g)$ have a negative
power of $(t^*-t)$ each. Other possible propagators emerging from diagrams
$h_1\,$--$\,h_{10}$ after the addition of a gluon going through the cut are
${\cal O}(1)$ in the limit $t^* \to 1$, $t^*-t \to 0$.
The overall scaling at leading order in $(1-t^*)$ results as
\begin{align}
\label{eq:realgluon}
 \int_{y}^1 d t^* \, (1-t^*)^{1/2-\eps-n_{t^*}}\int_x^{t^*} d t \, 
  (t^*-t)^{1-2\eps-n_{b^*}-n_{\bar t^*}}(1-t)^{-n_t}\, f(t) \,,
\end{align}
where $n_{t^*}$, $n_{\bar{t}^*}$ and $n_t$ are the numbers of top propagators with
momenta $(p_t + p_g)$, $(p_{\bar{t}} + p_g)$ and $p_t$, respectively,
$n_{b^*}$ stands for the number of bottom propagators with momentum $(p_b +
p_g)$, and $f(t)$ is a function of $\mathcal{O}(1)$ for $t \to 1$.
Two different regions are identified for the inner $t$-integration: the hard region
where $(t^*-t)\sim 1$, and the ultrasoft region where $(t^*-t)\sim (1-t^*)
\ll 1$.
After performing the integral  over $t$ in the hard region and in the ultrasoft region, 
we obtain contributions with scalings $(1-t^*)^{1/2-\eps-n_{t^*}}$ and
$(1-t^*)^{5/2-3\eps-n_{t^*}-n_{b^*}-n_{\bar t^*}-n_t}$, respectively, under the
$t^*$-integral.
Since the $n$'s are integer quantities, the endpoint-singular cases require either 
$n_{t^*}\ge 2$ or $n_{t^*}+n_{b^*}+n_{\bar t^*}+n_t \ge 4$. It is straightforward to check 
that these conditions are only accomplished by the real-gluon corrections to 
the NLO diagram $h_1$, which consist of three squared $e^+e^-\to bW^+ \bar{t}g$
diagrams and three interference diagrams, as shown in Fig.~\ref{fig3}. Moreover, the 
amplitudes from Fig.~\ref{fig3} involve only endpoint-singular cases of the
form $\int_y^1 d t^* \,
(1-t^*)^{-3/2-n\eps}$, producing contributions proportional to
$m_t/\Lambda$ according to~(\ref{eq:intep32}). Subleading terms in the expansion
in $(1-t^*)$ do not yield half-integer powers, so no $1/\eps$ endpoint singularities
arise from the real-gluon emission diagrams that contribute to the NNLO non-resonant
cross section.

\subsection{Calculation of endpoint singularities}
\label{sec:calcmethod}

The endpoint-singular contributions contained in the 
amplitudes of Figs.~\ref{fig2} and \ref{fig3} arise from
terms in the integrand of the outer phase-space integration
of the form $(1-t)^{r+n\epsilon}$, with $r=-2,-3/2,-1$ for the virtual-gluon corrections,
and of the form $(1-t^*)^{-3/2+n\epsilon}$ for the case of real-gluon emission. 
The extraction of such terms is greatly simplified if the 
1-loop and phase-space integrals over other variables are 
asymptotically expanded in $(1-t)$ and $(1-t^*)$ prior to their evaluation
using the strategy of expansion by
regions~\cite{Beneke:1997zp,Smirnov:2002pj,Jantzen:2011nz}.
The relevant regions are
determined and their completeness is confirmed by asymptotically expanding
Mellin--Barnes representations of the full integrals.
Such an interplay between expansion by regions and Mellin--Barnes
representations has been described in detail in~\cite{Jantzen:2011nz}.
For the 3-particle (virtual-gluon) diagrams, three regions are found in the loop
integration over the (gluon) momentum $k$:
the \emph{hard region}
where $k \sim m_t$, the \emph{ultrasoft region} where $k \sim m_t \,
(1-t)$, and the \emph{potential region} where $k^0 \sim m_t \, (1-t)$ and
$\vec k \sim m_t \, (1-t)^{1/2}$ (defined in the rest frame of~$p_t$).
Note that these regions are analogous to the ones that are found in the threshold
expansion~\cite{Beneke:1997zp} if we identify $v^2$ there with $(1-t)$.
For the 4-particle (real-gluon) diagrams, only the hard region with $k
\sim m_t \, (t^*-t) \sim m_t$ and the ultrasoft region with $k \sim m_t
\, (t^*-t) \sim m_t \, (1-t^*)$ contribute to the phase-space integral over
the variable~$t$. It is illustrative to show how the method of regions applies to
our case by means of two non-trivial examples.

Consider first the virtual gluon connecting the antitop with the on-shell
bottom quark, diagram $h_{1b}$ of Fig.~\ref{fig2}. The relevant 1-loop 
scalar integral reads 
\begin{align}
 I_{1b} &=
  \int d^dk  \, \frac{1}{((p_t+k)^2-m_t^2)\,((p_{\bar t}-k)^2-m_t^2)\,(p_b+k)^2\, k^2}
\nonumber
\\
 \label{eq:I1b}
 &=  \int d^dk  \, \frac{1}{(2p_t\cdot k + k^2 - m_t^2(1-t))\,( -2 p_{\bar t}\cdot k + k^2)\,(2p_b\cdot k+ k^2)\, k^2}\;,
\end{align}
where we have used the on-shell conditions for the antitop ($p_{\bar t}^2=m_t^2$)
and the bottom quark ($p_b^2=0$). In order to expand the denominators for each
region we need to know the scaling in $(1-t)$ of the
final-state particle momenta. Working in the top rest frame, where $p_t = (p_t^0,\vec{0})$,
simple kinematics gives, at leading order in $(1-t)$,
\begin{align}
 p_t^0 \simeq m_t  \;\;,\;\;
 p_{\bar t}^0 \simeq m_t    \;\;,\;\;  |\vec{p}_{\bar t}| \simeq \sqrt{2} \, m_t  (1-t)^{1/2} \;\;,\;\;
 p_b^0 = |\vec{p}_b| \simeq \frac{m_t}{2}\,(1-x) \;.
 \label{eq:kin}
\end{align}
Note that we regard neither $(1-x)$ nor $x$ as small quantities for the purpose of the $(1-t)$-expansion.
In the hard region where $k\sim m_t$, all the propagators scale as order $(1-t)^0$, 
and $I_{1b}^{(\rm h)}\sim  m_t^{-4-2\epsilon} (1-t)^0$. Taking into account the factor $(1-t)^{1/2-\epsilon}$
from the $q\to p_t + p_{\bar t}$ phase-space measure and $(1-t)^{-1}$ from the top propagator on the 
r.h.s.\ of the cut, the hard-region contribution scales as $(1-t)^{-1/2-\epsilon}$, and is thus endpoint-regular. 
In the ultrasoft region, we have to consider $k\sim m_t(1-t)$, which allows to drop $k^2$ from the fermion propagators
as well as $\vec{p}_{\bar t}\cdot\vec{k}$, obtaining
\begin{align}
 I_{1b}^{(\rm us)} = 
   \int d^dk  \, \frac{1}{( 2m_t k^0-m_t^2(1-t))\,( -2m_t k^0)\,(2p_{b,0}\cdot k)\, k^2} \,,
\label{eq:I1bus}
\end{align}
where $p_{X,0}$ denotes the first term in the $t\to 1$ expansion of the corresponding momentum $p_X$. Therefore 
we get that $I_{1b}^{(\rm us)}\sim  m_t^{-4-2\epsilon} (1-t)^{-1-2\epsilon}$,
and an overall endpoint-singular contribution to diagram $h_{1b}$ from the
ultrasoft contribution scaling as $(1-t)^{-3/2-3\epsilon}$. An additional
factor $\vec{p}_{\bar t}\cdot\vec{k} / (m_t k^0) \sim (1-t)^{1/2}$ in the
integrand of $I_{1b}^{(\rm us)}$
(coming, for instance, from the next-to-leading
term in the expansion of the antitop propagator in the ultrasoft region)
could potentially yield a $1/\epsilon_\ep$ endpoint divergence. However, the resulting
angular integral can be shown to vanish, so there are no $1/\epsilon_\ep$
endpoint divergences from the ultrasoft region.

Finally, the expansion of $I_{1b}$ in the potential region where $k^0 \sim \vec k^2/ m_t \sim (1-t)$ yields at leading order
\begin{align}
 I_{1b}^{(\rm p)} =
   \int d^dk  \, \frac{1}{(-\vec{k}^2+ 2m_t k^0-m_t^2(1-t))\,
   (-\vec{k}^2-2m_t k^0+2\vec{p}_{\bar t,0}\cdot\vec{k})\,(-2\vec{p}_{b,0}\cdot \vec{k})\,(-\vec{k}^2)} \,.
\label{eq:I1bpot}
\end{align}
{}From (\ref{eq:I1bpot}) we find $I_{1b}^{(\rm p)}\sim  m_t^{-4-2\epsilon} (1-t)^{-1-\epsilon}$, and thus a
leading-order scaling \mbox{$(1-t)^{-3/2-2\epsilon}$} for the potential-region contribution to diagram $h_{1b}$. 
Subleading terms of order\linebreak
\mbox{$(1-t)^{-1-2\epsilon}$} arise from the next-to-leading term in the expansion
of the bottom propagator (suppressed by $(1-t)^{1/2}$) as well as from an
additional factor of either $\vec{p}_{\bar t,0}\cdot\vec{p}_{b,0}$ or
$\vec{p}_{b,0}\cdot\vec{k}$ (both $\sim m_t^2(1-t)^{1/2}$) in the 
numerator of the full $h_{1b}$ amplitude. These subleading terms generate the
$1/\epsilon_\ep$ endpoint divergences from diagram $h_{1b}$.

Let us now consider an example involving a real-gluon correction, namely diagram $g_5$ from Fig.~\ref{fig3}.
In the rest frames of either $p_t$ or $p_{t^*}$, the components of the on-shell
gluon momentum are proportional to the variable $(t^*-t)$,
defined in Sec.~\ref{sec:nonresonantdivs},
such that both the bottom propagator with momentum $(p_b+p_g)$ and the
antitop propagator with momentum $(p_{\bar t}+p_g)$ provide a negative
power of $(t^*-t)$ each. So, according to (\ref{eq:realgluon}) with
$n_{t^*} = n_{\bar{t}^*} = n_t = n_{b^*} = 1$, the
integral over the variable $t$ has the form
\begin{align}
\label{eq:g5}
I_{g_5} = \int_x^{t^*} d t \, \frac{f(t)}{(1-t)(t^*-t)^{1+2\eps}}
  = \int_0^\infty du \, 
    \frac{\theta(t^*-x-u) \, f(t^*-u)}{(1-t^*+u) \, u^{1+2\eps}}\,,
\end{align}
where $f(t) \sim \Oc(1)$ for $t \to 1$, and we have introduced $u\equiv (t^*-t)$.
Applying the expansion by regions to integrals with finite boundaries or
theta functions is explained in~\cite{Jantzen:2011nz}.
We are interested in the expansion of this integral in $(1-t^*)$. As can easily
be confirmed by writing a Mellin--Barnes representation of the integral, for $t^*\to 1$ 
it receives contributions from the hard region, $u\sim 1\gg 1-t^*$, and from the
ultrasoft region, $u\sim 1-t^*$. In the former case we can expand out factors of $(1-t^*)$ in the 
denominator of (\ref{eq:g5}) and set $t^*\simeq 1$ in the numerator for the leading-order
contribution: 
\begin{align}
\label{eq:g5h}
I_{g_5}^{\rm (h)} = \int_0^{1-x} du \, 
 \frac{f(1-u)}{u^{2+2\eps}}\sim (1-t^*)^0 \,. 
\end{align}
So the overall scaling of the hard contribution is given by the remaining
factor in~(\ref{eq:realgluon}), $(1-t^*)^{-1/2-\eps}$, which is endpoint-regular.
However, in the ultrasoft region we have to set $\theta(t^*-x-u)\simeq
\theta(1-x) =1$ and $f(t^*-u) \simeq f(1)$ and
keep the first propagator unexpanded. Thus 
\begin{align}
\label{eq:g5us}
I_{g_5}^{\rm (us)} = f(1) \int_0^\infty du \, 
 \frac{1}{(1-t^*+u)\,u^{1+2\eps}}\sim (1-t^*)^{-1-2\eps} \,,
\end{align}
which provides the scaling $(1-t^*)^{-3/2-3\eps}$ for the 
leading ultrasoft term. By explicit computation it can be shown that 
no $(1-t^*)^{-1-3\eps}$ contribution is generated by subleading
terms in the expansion of the ultrasoft region because they involve
angular integrations of the form  $\int_{-1}^{+1} d\cos \theta (1-\cos^2 \theta)^{- \eps}\,\cos\theta$
which vanish.

\section{Results for the NNLO non-resonant contributions in the \boldmath$\Lambda/m_t$ expansion}
\label{sec:results}

The following results contain all endpoint-singular contributions to the
individual diagrams in Figs.~\ref{fig2} and~\ref{fig3}, {\it i.e.}, within the asymptotic
expansion described above,
the terms of~(\ref{eq:endpointsing}) with $r \ge 1$. Endpoint-regular
contributions with $r < 1$ are omitted. From the viewpoint of the results,
all terms of the expansion in powers of $\Lambda/m_t$ are provided up to and
including the order $(\Lambda/m_t)^0\ln (m_t^2/\Lambda^2)$, as shown in 
Eqs.\ (\ref{eq:intep2})--(\ref{eq:intep1}). The omitted terms are of
order $(\Lambda/m_t)^0$ without endpoint-singular $1/\eps$ terms or
logarithms of~$\Lambda$, or of higher order
in $(\Lambda/m_t)$; they are regular in the limit $\Lambda \to 0$.
The results presented here get dominant in the limit $\Lambda \to 0$, where
they are \emph{a priori} not valid because our effective-field-theory
treatment requires $\Lambda^2 \gg m_t \Gamma_t$ (see~\cite{Beneke:2010mp}).
However, our result taking $\Lambda$ as a tiny ($\Lambda \ll m_t$) unphysical scale
provides a precise approximation of the amplitude at the endpoint, which 
could be complemented by a numerical evaluation of the
contributions outside this tight $\Lambda$-cut (where the top
propagators never go on shell) in order get a precise evaluation of the 
full NNLO non-resonant corrections to the total cross section with or without invariant-mass
cuts.

All the diagrams relevant to this NNLO analysis with the exception of $h_{4a}$
have the structure of a hadronic tensor $H^{\mu\nu}$ connected to a leptonic
tensor via two photon
or $Z$ propagators. While the leptonic tensor contains the $e^+ e^-$ pairs on
the left-hand and right-hand sides of the forward-scattering amplitude, the
hadronic tensor involves everything between the two photon/$Z$ propagators,
including the cut through $b W^+ \bar t (g)$, with the exception of two
powers of the elementary charge~$e$ removed from the two external
vertices. In the case of photon/$Z$ couplings to quarks, these two vertices
on the left-hand and right-hand side of the hadronic tensor read
\begin{equation}
  \label{eq:Hcouplings}
  i \gamma^\mu \, (v_f^\tL - a_f^\tL \gamma_5)
  \quad \text{and} \quad
  i \gamma^\nu \, (v_f^\tR - a_f^\tR \gamma_5)
  \,,
\end{equation}
respectively, where $v_f^{\tL,\tR}$ and $a_f^{\tL,\tR}$ are the vector and
axial-vector couplings of the corresponding quark ($f=t,b$).
In the case of a $Z$-boson, they are given by 
\begin{eqnarray}
\label{eq:vfaf}
&& v_f^{\tL,\tR} =\frac{T_3^f-2Q_f s_w^2}{2s_w c_w},
\quad\qquad
a_f^{\tL,\tR}=\frac{T_3^f}{2s_w c_w},
\end{eqnarray}
where $s_w$ ($c_w$) is the sine (cosine) of the weak mixing angle,
$Q_f$ the electric charge of the fermion ($Q_t = 2/3$, $Q_b = -1/3$;
similarly $Q_e = -1$), and $T_3^f$ the third component of
the weak isospin of the fermion.
For a photon, $v_f^{\tL,\tR} = -Q_f$ and $a_f^{\tL,\tR} = 0$. 
In the case of diagram $h_{3a}$
the relevant 3-gauge-boson coupling is $I_{WW}^\tL$, multiplying the usual
momentum-dependent terms as defined in (\ref{eq:IWWL}) below.
We have $I_{WW}^\tL = 1$ for a photon attached to the left of the hadronic tensor
and $I_{WW}^\tL = -c_w/s_w$ for a $Z$-boson.

We note
that for all diagrams except for $g_i,\,i=1,2,3$, (which are already symmetric)
there is a symmetric contribution where the right and left parts of
the diagram have been exchanged. For the diagrams with the hadronic-tensor
structure (all except $h_{4a}$), the corresponding amplitude for the symmetric amplitude
is obtained by replacing the coupling factors on the sides of the hadronic tensor as
\begin{equation}
  \label{eq:XLYR}
  X^\tL Y^\tR \to Y^\tL X^\tR \,,
\end{equation}
where $X,Y = v_t, a_t, v_b, a_b, I_{WW}$, and then taking 
the complex conjugate of the whole amplitude. The results in this section correspond
to the amplitudes with $b W^+ \bar{t} (g)$ cuts, as shown in
Figs. \ref{fig2} and~\ref{fig3}. The 
additional diagrams with $\bar{b} W^- t (g)$ cuts also have to be taken into account when computing
the non-resonant contribution to $e^+e^-\to W^+W^- b\bar{b}$, but their contribution
is equal to the corresponding one with $b W^+ \bar t (g)$ cuts by virtue of 
$CP$-invariance.

The hadronic tensor has the Lorentz structure $H^{\mu\nu} = H \, g^{\mu\nu}
+ \tilde H \, q^\mu q^\nu/q^2$ with the total momentum $q = p_{e^+} +
p_{e^-} = p_b + p_W + p_{\bar t} \, (+ p_g)$ and $q^2 = s$. Only the first
term contributes to the cross section, it is projected out by
\begin{equation}
  H = \frac{1}{3-2\eps} \left( g_{\mu\nu} - \frac{q_\mu q_\nu}{q^2} \right) H^{\mu\nu}
  \,.
\end{equation}

For the propagators and polarization sums of gluons, Feynman gauge is used,
while unitarity gauge is employed for $W$-bosons. All  NNLO contributions
are proportional to the factor
\begin{align}
 \label{eq:Hnorm}
 N_{\eps}  &= \left(\frac{\mu^2}{m_t^2}\right)^{3\eps}  m_t \Gamma_t^{\rm Born} \, \Nc \CF \,
    \frac{\alpha_s}{4\pi}   \,,
\end{align}
where $\mu$ is the scale introduced in
dimensional regularization and $\Nc=3$ is the number of QCD colours.
$\Gamma_t^{\rm Born}$ is the tree-level top decay width,
\begin{equation}
\Gamma_t^{\rm Born}=\frac{\alpha|V_{tb}|^2 m_t}{16s_w^2} 
\frac{(1-x)^2(1+2x)}{x}\,, 
\label{eq:Gammatop}
\end{equation}
obtained
from the amplitude $t\to b W^+$ with the bottom-quark mass set to zero.
In the presentation of the results that follows, 
we shall mark if the $1/\eps$ poles are of ultraviolet,
infrared or endpoint-singular origin by writing $1/\eps_\UV$, $1/\eps_\IR$ or
$1/\eps_\ep$, respectively. 

\subsection{Virtual corrections to diagram \boldmath $h_1$}

We first give results for the hadronic tensor coefficient $H$
from the virtual ${\cal O}(\alpha_s)$ corrections
to diagram $h_1$, \textit{i.e.} diagrams $h_{1X},\,X=a,\dots g$, depicted in Fig.~\ref{fig2},
that generate the three types of endpoint-divergent contributions listed in (\ref{eq:intep2})--(\ref{eq:intep1}).
It is a general feature of these corrections that 
$(m_t^2/\Lambda^2)$ and $(\Lambda/m_t)^0\ln (m_t^2/\Lambda^2)$ terms 
arise only from the potential region in the 1-loop integration over the gluon
momenta, whereas 
the $(m_t/\Lambda)$ terms come from either hard or ultrasoft gluons (with the only 
exception of diagram $h_{1b}$, as explained below). 

The amplitude for $h_{1a}$ arises from inserting the virtual-gluon correction in
the $t\bar{t}$-vertex and has the highest degree of endpoint singularity, as 
explained in Sec.~\ref{sec:nonresonantdivs}. The result reads
\begin{align}
 \label{eq:H1a}
  H_{1a}
   = N_\eps \,
   & \biggl\{ 
      2\,\frac{m_t^2}{\Lambda^2} \, v_t^\tL v_t^\tR 
      + \frac{m_t}{\Lambda} \, v_t^\tL v_t^\tR
         \frac{\sqrt{2}}{\pi^2} \left[
          \frac{1}{\eps_\UV} - 3\ln\frac{m_t^2}{\Lambda^2} - 2\ln(1-x) - \ln 2
          + \frac{2 \, (1+x)}{1+2x} \right]
\nonumber \\ & \;
      + \left( \frac{1}{\eps_\ep} + 2  \ln\frac{m_t^2}{\Lambda^2} \right) 
        \left[ - v_t^\tL v_t^\tR \,  \frac{2 \, (2+2x+5x^2)}{3 \, (1-x) \, (1+2x)}
        - \frac{1}{4} v_t^\tL a_t^\tR 
        + \frac{1}{6} a_t^\tL a_t^\tR  \right] \,
    \biggr\}
  \,.
\end{align}
Terms of order $(\Lambda/m_t)^0$ without $1/\eps_\ep$ terms or logarithms of~$\Lambda$ and
endpoint-regular contributions of $\Oc(\Lambda/m_t)$ are omitted in our
results, as explained above. Also terms of $\Oc(\eps)$ are dropped
everywhere.

For the virtual gluon connecting the bottom and the antitop quark, diagram $h_{1b}$, the 
hadronic tensor coefficient reads 
\begin{align}
 \label{eq:H1b}
  H_{1b}
   = N_\eps 
     & \biggl\{ 
       \frac{m_t}{\Lambda} \, v_t^\tL v_t^\tR
         \frac{\sqrt{2}}{\pi^2} \left[ \,
          -\frac{1}{\eps_\IR^2}
          + \frac{1}{\eps_\IR} \left(
            -3 \ln\frac{m_t^2}{\Lambda^2} + 2 \ln(1-x) + \ln 2 +
            \frac{2\,(1+3x)}{1+2x} \right)
          \right.
\nonumber \\* & \qquad \qquad \quad \;\;\;\;
          - \frac{9}{2} \ln^2\frac{m_t^2}{\Lambda^2}
          + 3 \left( 2 \ln(1-x) + \ln 2 + \frac{2\,(1+3x)}{1+2x} \right)
            \ln\frac{m_t^2}{\Lambda^2}
\nonumber \\ & \qquad \qquad \quad \;\;\;\;
          - 2 \ln^2(1-x)
          - 2 \left( \ln 2 + \frac{2\,(1+3x)}{1+2x} \right) \ln(1-x)
          - \frac{1}{2} \ln^2 2
\nonumber \\ & \qquad \qquad \quad \;\;\;\;
          - \frac{2\,(1+3x)}{1+2x} \ln 2
          - \frac{4\,(6+13x)}{1+2x}
          + \frac{\pi^2}{12}
\nonumber \\ & \qquad \qquad \quad \;\;\;\;
          \left. + \, i \pi \left( -\frac{2}{\eps_\IR} -4\ln\frac{m_t^2}{\Lambda^2} + 4\ln(1-x) 
            + \frac{4 \, (1+3x)}{1+2x}  \right) \right]
\nonumber \\ & \;
       + \left( \frac{1}{\eps_\ep} + 2  \ln\frac{m_t^2}{\Lambda^2} \right) 
        \left[  v_t^\tL v_t^\tR \,  \frac{3 \, (1+x+2x^2)}{4 \, (1-x) \, (1+2x)}
        - ( v_t^\tL a_t^\tR - a_t^\tL v_t^\tR )\frac{1-2x}{6(1+2x)} \right] \,
    \biggr\}
  \,.
\end{align}
The single and double infrared singularities in (\ref{eq:H1b}) are related to the emission of the
virtual gluon from the massless bottom quark. The contribution proportional
to $i\pi$ of order $m_t/\Lambda$ in the second-to-last line of (\ref{eq:H1b})
arises from potential-gluon momentum,
but, as it is purely imaginary, it cancels with the symmetric contribution
where the gluon is exchanged on the r.h.s. of the cut.

The virtual gluon correcting the $tb$-vertex, diagram $h_{1c}$, gives
\begin{align}
 \label{eq:H1c}
  H_{1c}
   = N_\eps \,
       \frac{m_t}{\Lambda} \, v_t^\tL v_t^\tR
        \frac{2\sqrt{2}}{\pi^2}  & \biggl[ \,
         \frac{1}{2 \, \eps_\UV} + \frac{1}{\eps_\IR} \left(
          \ln\frac{m_t^2}{\Lambda^2} + \ln(1-x) - 3 \right)
\nonumber \\* &
         + 2 \ln^2\frac{m_t^2}{\Lambda^2}
         - \left( \ln(1-x) + \ln 2 + \frac{9+22x}{2 \, (1+2x)} \right)
         \ln\frac{m_t^2}{\Lambda^2}
\nonumber \\ &
          - 3\ln^2(1-x) + \left( \frac{7+15x}{1+2x} - \ln 2 \right) \ln(1-x)
         + \frac{5}{2} \ln 2 
\nonumber \\* &
         + \frac{5 \, (1+3x)}{1+2x}  - \Li2(x) + \frac{\pi^2}{6} \,
        \biggr] \,.
\end{align}
The remaining virtual corrections to diagram~$h_1$ correspond to
renormalization and self-energy contributions.
At ${\cal O}(\alpha_s)$, the top- and bottom-quark fields and the top mass need to be
renormalized, which is done in the on-shell scheme. Since the  non-resonant NLO diagrams
are purely of electroweak origin, they do not involve QCD couplings, so the renormalization
of $\alpha_s$ is irrelevant for the current NNLO analysis. The insertion of the renormalized
self-energy into a top/antitop line with momentum $p$ next to a
cut vanishes, since this contribution is proportional to
$(m_t^2-p^2)^{-2\eps}$ in the on-shell limit, and the Cutkosky rules
prescribe a factor $\delta(p^2-m^2)$ which sets $(m_t^2-p^2)^{-2\eps}$ to
zero as a scaleless term in dimensional regularization. The same is true for
a cut bottom line, because the bottom self-energy is scaleless and
vanishes for $p_b^2=0$. Therefore we do not need
to consider self-energy insertions in lines which are cut. On the other hand,
when the top self-energy is inserted
into an internal top line, the top-field renormalization parts of the
vertex counterterms from the two adjacent vertices cancel
exactly the contribution of the field-renormalization to the 2-point
top counterterm in the renormalized self-energy
insertion, so the complete correction to an internal top line is equal to the insertion
of the bare self-energy plus the mass-renormalization
part of the 2-point top counterterm. This correction only generates 
an endpoint-singular contribution when inserted into the internal top lines 
of the NLO diagram $h_1$. The corresponding NNLO diagrams are $h_{1d}$ and $h_{1e}$ in Fig.~\ref{fig2},
and the sum of both gives the coefficient 
\begin{align}
 \label{eq:H1de}
  H_{1de}
     = N_\eps \,
       \frac{m_t}{\Lambda} \, v_t^\tL v_t^\tR
       \frac{\sqrt{2}}{\pi^2}  & 
       \biggl[
        -\frac{1}{\eps_\UV}
        + 3\ln\frac{m_t^2}{\Lambda^2}
        + 2\ln(1-x)
        + \ln 2
        - \frac{2 \, (5+9x)}{1+2x} 
     \biggr]\,.
\end{align}
The bare self-energy entering diagram $h_{1d}$ receives contributions from 
hard and ultrasoft loop momenta, whereas the mass- as well as the field-renormalization
constants are entirely determined by hard contributions. 
For the computation of $h_{1e}$ (and also for $h_{1f}$ and $h_{1g}$) we need the result
for the NLO amplitude $h_1$ retaining ${\cal O}(\eps)$ terms, which was not
necessary for the calculation performed in \cite{Beneke:2010mp}. For completeness we write here the 
hadronic tensor coefficient for the endpoint-singular term of $h_1$ with
the full $\eps$-dependence:
\begin{align}
  \label{eq:H1}
  H_1^{\rm NLO} &= \left(\frac{\mu^2}{m_t^2}\right)^{2\eps}
    m_t \Gamma_t^{\rm Born} \, \Nc \,
    v_t^\tL v_t^\tR \, \frac{m_t}{\Lambda} \,
    \frac{2^{1/2-\eps} \, e^{2\eps\gamma_E} \, \Gamma^2(1-\eps)}{
    \pi^2 \, (1+2\eps) \, \Gamma^2(2-2\eps)} \,
    \frac{1+2(1-\eps)x}{(1+2x) \, (1-x)^{2\eps}}
    \left(\frac{m_t^2}{\Lambda^2}\right)^\eps
  \,.
\end{align}
Although the renormalized self-energy insertion into a cut bottom line
vanishes, there is a contribution from the bottom-field renormalization 
part $\delta_b/2$ of the counterterm of the $tbW$ vertex, diagram $h_{1f}$. 
This correction is obtained by multiplying the NLO result
$H_1^{\rm NLO}$~(\ref{eq:H1}) with $\delta_b/2$, where $\delta_b$ is defined 
from the relation between the bare and renormalized bottom-quark field, 
$b_0=(1+\delta_b)^{1/2}b$, 
\begin{equation}
  \label{eq:deltab}
  \delta_b =
    \left(\frac{\mu^2}{m_t^2}\right)^\eps \, \CF \, \frac{\alpha_s}{4\pi} \left(
      -\frac{1}{\eps_\UV} + \frac{1}{\eps_\IR} \right)
    + \Oc(\alpha_s^2)
    \,,
\end{equation}
obtaining
\begin{align}
  \label{eq:H1f}
  H_{1f} =
    N_\eps \,
    \frac{m_t}{\Lambda} \, v_t^\tL v_t^\tR \,
    \frac{\sqrt{2}}{\pi^2}
    \left( -\frac{1}{2\,\eps_\UV} + \frac{1}{2\,\eps_\IR} \right)\,,
\end{align}
which is effectively zero, but exhibits separate ultraviolet and infrared
singularities.

Finally, there is a contribution from top-field renormalization, diagram $h_{1g}$,
attributed to the antitop part $\delta_t/2$ of the counterterm of the
$t\bar{t}$-vertex. The top-field counterterm $\delta_t$, defined analogously to $\delta_b$, is determined
from the derivative of the bare self-energy with respect to $\dslash p_t$, giving
\begin{align}
  \label{eq:deltat}
  \delta_t &=
    \left(\frac{\mu^2}{m_t^2}\right)^\eps \, \CF \, \frac{\alpha_s}{4\pi}
    \left( -\frac{1}{\eps_\UV} - \frac{2}{\eps_\IR} - 4 \right)
    + \Oc(\alpha_s^2)
    \,.
\end{align}
Multiplying $\delta_t/2$ with (\ref{eq:H1}) we get for the hadronic tensor coefficient 
from diagram $h_{1g}$:
\begin{align}
  \label{eq:H1g}
  H_{1g} =
    N_\eps \,
    \frac{m_t}{\Lambda} \, v_t^\tL v_t^\tR \,
    \frac{\sqrt{2}}{\pi^2} \,
    & \biggl[
      -\frac{1}{2\,\eps_\UV} - \frac{1}{\eps_\IR}
      - \frac{3}{2} \ln\frac{m_t^2}{\Lambda^2}
      + 3\ln(1-x)
      + \frac{3}{2} \ln 2
      - \frac{5+7x}{1+2x} \,
    \biggr] \,.
\end{align}
Let us note that in the sum of all virtual-gluon corrections and renormalization 
contributions to the diagram $h_1$ listed above, the ultraviolet
$1/\eps_\UV$ singularities cancel out, such that the overall ${\cal O}(\alpha_s)$
correction is ultraviolet-finite. (Recall that the $1/\eps$ divergences in the
real-gluon corrections discussed next can only be of infrared origin.)

\subsection{Real-gluon corrections}
\label{subsec:realgluon}

As explained in Sec.~\ref{sec:nonresonantdivs}, for endpoint-singular contributions involving real gluons
we need to consider only the forward-scattering amplitudes with a 4-particle cut  corresponding
to real-gluon corrections to the NLO diagram $h_1$. These have been shown explicitly in Fig.~\ref{fig3} 
for the case of $bW^+\bar{t}g$ cuts. The real-gluon corrections involve three squared
$e^+e^- \to bW^+\bar{t}g$ amplitudes (diagrams $g_1, g_2, g_3$) and three interference
amplitudes (diagrams $g_4, g_5, g_6$) that have a symmetric counterpart which is obtained by
mirroring the diagram across the cut (without reversing the fermion flow). 

The contributions to the hadronic tensor from the symmetric diagrams read
\begin{align}
  \label{eq:Hg1}
  H_{g_1} =
    N_\eps \,\frac{m_t}{\Lambda} \, v_t^\tL v_t^\tR \,
    \frac{\sqrt{2}}{\pi^2} &
    \biggl[
      -4 \left( \ln\frac{m_t^2}{\Lambda^2} + \ln(1-x) \right)
      + \frac{x^2 \, (3+10x)}{(1-x)^2 \, (1+2x)} \ln x
\nonumber\\ & \;
      + \frac{109+85x-116x^2}{6 \, (1-x) \, (1+2x)}
    \biggr] \,,
\end{align}
\begin{align}
  \label{eq:Hg2}
  H_{g_2} =
    N_\eps \,\frac{m_t}{\Lambda} \, v_t^\tL v_t^\tR \,
    \frac{\sqrt{2}}{\pi^2} &
    \biggl[
      -\frac{1}{\eps_\IR}
      - \ln\frac{m_t^2}{\Lambda^2} + 4\ln(1-x) + \ln 2
      + \frac{x^2 \, (3-2x)}{(1-x)^2 \, (1+2x)} \ln x
\nonumber\\ & \;
      - \frac{29+5x-40x^2}{6 \, (1-x) \, (1+2x)}
    \biggr]\,,
\end{align}
\begin{align}
  \label{eq:Hg3}
  H_{g_3} =
    N_\eps \,\frac{m_t}{\Lambda} \, v_t^\tL v_t^\tR \,
    \frac{2 \sqrt{2}}{\pi^2} &
    \biggl[
      \frac{1}{\eps_\IR}
        + 3 \ln\frac{m_t^2}{\Lambda^2} - 2\ln(1-x) - \ln 2
        + \frac{2 \, (1+x)}{1+2x} 
    \biggr]\,.
\end{align}
While diagram $g_1$ receives contributions from both hard and ultrasoft gluons, 
diagram~$g_2$ only gets contributions from the hard region and
diagram~$g_3$ only from the ultrasoft region.

In the case of the interference diagrams, the contributions of both
$g_4$ and $g_5$ are entirely produced by ultrasoft gluons, whereas $g_6$
receives as well contributions from hard gluons. 
The results for the interference diagrams read
\begin{align}
  \label{eq:Hg4}
  H_{g_4} =
    -N_\eps \,\frac{m_t}{\Lambda} \, v_t^\tL v_t^\tR \,
    \frac{4\sqrt{2}}{\pi^2}\,,
\end{align}
\begin{align}
  \label{eq:Hg5}
  H_{g_5} =
    N_\eps \,\frac{m_t}{\Lambda} \, v_t^\tL v_t^\tR \,
    \frac{\sqrt{2}}{\pi^2} & \biggl[
      \frac{1}{\eps_\IR^2}
      + \frac{1}{\eps_\IR} \left(
        3 \ln\frac{m_t^2}{\Lambda^2} - 2 \ln(1-x) - \ln 2
        - \frac{2\,(1+3x)}{1+2x} \right)
\nonumber \\ & \;
      + \frac{9}{2} \ln^2\frac{m_t^2}{\Lambda^2}
      - 3 \left( 2 \ln(1-x) + \ln 2 + \frac{2\,(1+3x)}{1+2x} \right)
        \ln\frac{m_t^2}{\Lambda^2}
\nonumber \\ & \;
      + 2 \ln^2(1-x)
      + 2 \left( \ln 2 + \frac{2\,(1+3x)}{1+2x} \right) \ln(1-x)
      + \frac{1}{2} \ln^2 2 
\nonumber \\ & \;
      + \frac{2\,(1+3x)}{1+2x} \ln 2
      + \frac{4\,(6+13x)}{1+2x} - \frac{\pi^2}{12}
    \biggr]\,,
\end{align}
\begin{align}
  \label{eq:Hg6}
  H_{g_6} =
    N_\eps \,\frac{m_t}{\Lambda} \, v_t^\tL v_t^\tR \,
    \frac{2\sqrt{2}}{\pi^2} & \biggl[
      \frac{1}{\eps_\IR} \left(
        -\ln\frac{m_t^2}{\Lambda^2} - \ln(1-x) + 3 \right)
      - 2 \ln^2\frac{m_t^2}{\Lambda^2}
\nonumber \\ & \;
      + \left( \ln(1-x) + \ln 2 + \frac{5+12x}{1+2x} \right)
        \ln\frac{m_t^2}{\Lambda^2}
      + 3\ln^2(1-x)
\nonumber \\ & \;
      + \left( \ln 2 - \frac{2 \, (5+9x)}{1+2x} \right) \ln(1-x)
      - 3\ln 2
      - \frac{x \, (2+x)\ln x}{2\, (1-x)^2 \, (1+2x)} 
\nonumber \\ & \;
      - \frac{43+97x-122x^2}{12 \, (1-x) \, (1+2x)}
      + \Li2(1-x) - \frac{2\pi^2}{3}
      \biggr]\,.
\end{align}

It can be checked that in the sum of the real-gluon and virtual-gluon corrections to diagram
$h_1$, the infrared-singular
terms cancel out completely. The cancellation of infrared divergences holds independently
for the combinations $h_{1b}+g_5$, $h_{1c}+g_6$, $2 h_{1f}+ g_2$ and $2h_{1g}+ g_3$, where 
the factor $2$ in front of $h_{1f}$ and $h_{1g}$ accounts for the symmetric contribution. It is also
interesting to note that the sum of the
ultrasoft-gluon (real and virtual) contributions in the ${\cal O}(\alpha_s)$ corrections 
to diagram $h_1$ above vanishes. In particular, the ultrasoft pieces in the combinations
$h_{1a}+h_{1d}+h_{1e}$, $h_{1b}+g_5$, $h_{1c}+g_6$ and  $g_{1}+g_{3}+2g_{4}$,
cancel out separately.
Since there are no potential-region contributions of order $m_t/\Lambda$ and 
the remaining endpoint-singular diagrams $h_{ia},\; i=2,3,4$, give only 
$(\Lambda/m_t)^0\,\ln(m_t^2/\Lambda^2)$ terms, the order $m_t/\Lambda$ in the 
non-resonant cross section originates purely from the hard region.
This also implies that  logarithms $\ln(m_t^2/\Lambda^2)$
will be absent in the $m_t/\Lambda$ term: In
individual contributions, they arise from expanding hard-region terms
$\eps^{-k} \, (m_t^2/\Lambda^2)^{\eps}$ and ultrasoft-region
terms $\eps^{-k} \, (m_t^2/\Lambda^2)^{3\eps}$ in powers
of~$\eps$. Due to the different scaling of these two regions with
$m_t^2/\Lambda^2$, logarithms $\ln(m_t^2/\Lambda^2)$ may remain in a
contribution even if it is finite for $\eps \to 0$, cf.\ the amplitude~(\ref{eq:Hg1})
for diagram~$g_1$. But where the contributions from both
hard and ultrasoft regions are separately finite or, as here, the ultrasoft
contribution vanishes completely, leaving only an ultraviolet- and
infrared-finite hard contribution, there the logarithms
$\ln(m_t^2/\Lambda^2)$  must disappear.

The cancellation of ultrasoft terms in the NNLO non-resonant calculation is  
closely connected with the absence of contributions from 
ultrasoft gluons in the NNLO matrix element (\ref{eq:Ares}) containing
the interactions among the resonant top and antitop quarks. We postpone 
the discussion of this issue to Sec.~\ref{sec:psm}.

\subsection{Virtual corrections to diagrams \boldmath $h_2, h_3, h_4$}

The NNLO diagrams $h_{2a}$, $h_{3a}$ and $h_{4a}$ of Fig.~\ref{fig2}
produce endpoint-singular 
contributions of the form (\ref{eq:intep1}). The additional power $(1-t)^{-1/2-\eps}$
needed to produce the $1/\eps_\ep$ from the NLO diagrams $h_{2}$, $h_{3}$ and $h_{4}$ 
arises from the potential region of the $t\bar{t}$-vertex correction. 
The hadronic tensor coefficients from $h_{2a}$ and $h_{3a}$ read
\begin{align}
  \label{eq:H2a}
  H_{2a} & =
    N_\eps \, v_t^\tL \, (v_b^\tR + a_b^\tR) \, 
    \frac{1-5x-2x^2}{12 \, (1+x) \, (1+2x)} 
    \left( \frac{1}{\eps_\ep} + 2  \ln\frac{m_t^2}{\Lambda^2} \right) 
    \,,
  \\
  \label{eq:H3a}
  H_{3a} & =
    - N_\eps \, I_{WW}^\tL \, v_t^\tR \, 
    \frac{ 2+5x-2x^2}{12 x \, (1+2x)} 
    \left( \frac{1}{\eps_\ep} + 2  \ln\frac{m_t^2}{\Lambda^2} \right) 
    \,,
\end{align}
where the coupling factor $I_{WW}^\tL$ in (\ref{eq:H3a}) is defined from the 3-gauge-boson vertex
Feynman rule
\begin{equation}
 \label{eq:IWWL}
  i \, I_{WW}^\tL \Bigl[
    g^{\mu\rho} \, (q + p_{W^-})^\sigma
    + g^{\rho\sigma} \, (-p_{W^-} + p_{W^+})^\mu
    + g^{\sigma\mu} \, (-p_{W^+} - q)^\rho
  \Bigr]
  \,,
\end{equation} 
omitting the elementary charge~$e$. The total momentum $q = p_{W^+} +
p_{W^-}$ is incoming from the left-hand side (index~$\mu$), the lower
$W^-$-boson has outgoing momentum~$p_{W^-}$ (index~$\rho$), and the upper
$W^+$-boson has outgoing momentum~$p_{W^+}$ (index~$\sigma$). 

For diagram $h_{4a}$ we have to evaluate directly its contribution to 
the unpolarized $e^+ e^- \to b W^+ \bar t$ cross section, since it
does not have the structure of a hadronic tensor contracted
with a leptonic tensor. The cross-section contribution of 
diagram $h_{4a}$, already summed over the two gauge bosons,
photon and $Z$, in the $s$-channel propagator reads
\begin{align}
\label{eq:sigmah4a}
  \Delta\sigma_{4a} & =
    - N_\eps \,
    \frac{\pi^2 \, \alpha^2}{\sw^2} \,
    \frac{1}{s}
    \left( \frac{Q_t \, Q_e}{s} + \frac{v_t \, (v_e + a_e)}{s - M_Z^2} \right)
    \frac{x}{(1-x)^3 \, (1+2x)} \,
\nonumber \\* &\qquad\qquad \times
     \left[
        4 \ln\!\left(\frac{2}{x}-1\right)
        + \frac{(1-x) \, (1-2x-23x^2)}{3x^2}
     \right]
     \left( \frac{1}{\eps_\ep} + 2  \ln\frac{m_t^2}{\Lambda^2} \right)  \,,
\end{align}
where the squared centre-of-mass energy~$s$
has been set to $4m_t^2$ only in the non-trivial parts of the loop and
phase-space integrations, but kept general in the photon and $Z$ propagators and
in the kinematic factors of the cross section. This
corresponds to the same prescription used for the NLO non-resonant 
contributions computed in~\cite{Beneke:2010mp}.

\subsection{Complete endpoint-singular NNLO non-resonant cross section}
\label{sec:NNLOres}

With all the results from the individual diagrams of Figs.~\ref{fig2}
and \ref{fig3} at hand, we can compute the complete endpoint-singular
non-resonant NNLO contribution to the $e^+e^-\to  W^+W^-b\bar{b}$
cross section. 

First, we need to sum up the contributions which come in the
form of a hadronic tensor coefficient (all diagrams except $h_{4a}$).
The real-gluon contributions
$H_{g_1}$~(\ref{eq:Hg1}),
$H_{g_2}$~(\ref{eq:Hg2}) and
$H_{g_3}$~(\ref{eq:Hg3})
from symmetric diagrams have to be counted once.
For all other diagrams, there are corresponding symmetric diagrams, so the
virtual-gluon contributions
$H_{1a}$~(\ref{eq:H1a}),
$H_{1b}$~(\ref{eq:H1b}),
$H_{1c}$~(\ref{eq:H1c}),
$H_{2a}$~(\ref{eq:H2a}) and
$H_{3a}$~(\ref{eq:H3a}),
the renormalization and self-energy contributions
$H_{1de}$~(\ref{eq:H1de}),
$H_{1f}$~(\ref{eq:H1f}) and
$H_{1g}$~(\ref{eq:H1g}),
and the real-gluon interference contributions
$H_{g_4}$~(\ref{eq:Hg4}),
$H_{g_5}$~(\ref{eq:Hg5}) and
$H_{g_6}$~(\ref{eq:Hg6})
have to be symmetrized before adding them together. This is done by
replacing pairs of coupling factors as specified in (\ref{eq:XLYR}).
The contribution of each diagram (with hadronic tensor coefficient~$H$) to
the unpolarized $e^+ e^- \to b W^+ \bar t (g)$ cross section is then given by 
\begin{equation}
  \label{eq:DeltasigmaH}
  \Delta\sigma = - 8(1-\eps)\pi^2 \, \alpha^2 \,
    \frac{v_e^\tL v_e^\tR + a_e^\tL a_e^\tR}{(s-M_\tL^2) \, (s-M_\tR^2)} \,
    H \,,
\end{equation}
where $\alpha$ is the electromagnetic coupling and 
$M_\tL$ and $M_\tR$ are the masses of the gauge bosons
to the left-hand and right-hand sides of the hadronic tensor; the latter
are attached to the $e^+e^-$ initial state via couplings
$v_e^{\tL,\tR}$ and $a_e^{\tL,\tR}$,
defined analogously to~(\ref{eq:Hcouplings}).
The contribution (\ref{eq:DeltasigmaH}) has to be
summed over the four combinations of gauge bosons ($\tL,\tR =
\text{photon},Z$) and, in addition, it must be counted twice, because for every
contribution with $b W^+ \bar t (g)$ cut presented here, an equivalent one
with $\bar b W^- t (g)$ cut exists.

The result from the correction $\Delta\sigma_{4a}$~(\ref{eq:sigmah4a})
must be counted four times (symmetrization and counting both cuts
$b W^+ \bar t$ and $\bar b W^- t$). After adding this piece, the following
complete endpoint-singular non-resonant NNLO contribution to the cross section
results:

\begin{align}
  \label{eq:sigmatot}
  &\sigma_{\rm non-res}^{(2),\ep}
    = \frac{32\pi^2  \alpha^2}{s} \,
      \frac{\Gamma_t^{\rm Born}}{m_t} \, \Nc
\nonumber \\* &\quad{}\times
    \Biggl\{
      \Bigl[ Q_t^2 \, C_{\gamma\gamma}(s)
        - 2 Q_t v_t \, C_{\gamma Z}(s) + v_t^2 \, C_{ZZ}(s) \Bigr] \,
      \biggl\{
        2 \CF \, \frac{\alpha_s}{\pi} \, \, \frac{m_t^2}{\Lambda^2} \,
\nonumber \\ &\qquad{}\qquad{}\qquad{}\quad{}
        + \frac{2\sqrt{2}}{\pi^2} \, \frac{m_t}{\Lambda} \,
          \Big(
             \delta \Gamma_t^{(1)} - 4\CF \, \frac{\alpha_s}{\pi}
          \Big)
      \biggl\}
\nonumber \\ &\quad\quad\;\;
     + \biggl\{ - \Bigl[ Q_t^2 \, C_{\gamma\gamma}(s)
        - 2 Q_t v_t \, C_{\gamma Z}(s) + v_t^2 \, C_{ZZ}(s) \Bigr]\, \frac{7+7x+22x^2}{6(1-x) \, (1+2x)} 
\nonumber \\ &\quad\qquad\quad\;
     + \frac{1}{3} a_t^2 \, C_{ZZ}(s) 
          + \frac{1}{2} Q_t a_t \, C_{\gamma Z}(s) - \frac{1}{2}v_t a_t \, C_{ZZ}(s) 
\nonumber \\ &\quad\qquad\quad\;
      + \Bigl[ Q_t Q_b \, C_{\gamma\gamma}(s)
          - \bigl( Q_t \, (v_b+a_b) + Q_b v_t \bigr) \, C_{\gamma Z}(s)
          + v_t \, (v_b+a_b) \, C_{ZZ}(s) \Bigr]
\nonumber \\* &\qquad{}\qquad{}\qquad{}\quad{}\times
        \frac{1-5x-2x^2}{6(1+x) \, (1+2x)}
\nonumber \\ &\quad\qquad\quad\;
      + \left[ Q_t \, C_{\gamma\gamma}(s)
          - \left( v_t + Q_t \, \frac{\cw}{\sw} \right) C_{\gamma Z}(s)
          + v_t \, \frac{\cw}{\sw} \, C_{ZZ}(s) \right]
        \frac{2+5x-2x^2}{6x \, (1+2x)}        
\nonumber \\ &\quad\qquad\quad\;
      - \Bigl[ Q_t \, C_\gamma(s) + v_t \, C_Z(s) \Bigr] \,
        \left[
            \ln\!\left(\frac{2}{x}-1\right)
            + \frac{(1-x) \, (1-2x-23x^2)}{12x^2}
        \right]
\nonumber \\* &\qquad{}\qquad{}\qquad{}\quad{} \times
        \frac{x}{4(1-x)^3 \, (1+2x)} 
        \,\biggr\}\,
        \CF \, \frac{\alpha_s}{2\pi} \,\left(
           \frac{1}{\eps_\ep} + 2  \ln\frac{\mu_{\rm soft}^2}{\Lambda^2}
        \right)
    \Biggr\}
  \,,
\end{align}
where we have used the functions 
\begin{gather}
  C_{\gamma\gamma}(s) = -Q_e^2 \, \frac{m_t^2}{4s} \,,\qquad
  C_{\gamma Z}(s) = \frac{Q_e v_e \, m_t^2}{4 \, (s-M_Z^2)} \,,\qquad
  C_{ZZ}(s) = -\frac{(v_e^2 + a_e^2) \, m_t^2 \, s}{4 \, (s-M_Z^2)^2} \,,
\nonumber \\*
  C_\gamma(s) = \frac{Q_e \, m_t^2}{\sw^2 \, s} \,,\qquad
  C_Z(s) = \frac{(v_e + a_e) \, m_t^2}{\sw^2 \, (s-M_Z^2)} \,,
\end{gather}
which arise from the photon and $Z$-boson propagators, 
and dropped ${\cal O}(\epsilon)$ terms.
In order to improve the quality of the EFT expansion, we
have kept $q^2 = s$ general in all photon and $Z$ propagators as well as in the
kinematic factors of the cross section and set $s=4m_t^2$ only in the
non-trivial parts of the loop and phase-space integrations, as
done in~\cite{Beneke:2010mp}.
We comment below on the introduction of the scale~$\mu_{\rm soft}$
in~(\ref{eq:sigmatot}).

The coefficient $\delta \Gamma_t^{(1)}$ in the $m_t/\Lambda$ term
of~(\ref{eq:sigmatot}) is equal to
\begin{align}
 \label{eq:deltaGamma}
   \delta \Gamma_t^{(1)} &= \CF \, \frac{\alpha_s}{2\pi} \,
          \biggl[
            - \left( 2 \ln x + \frac{5+4x}{1+2x} \right) \ln(1-x)
            - 4 \,\Li2(x)
            - \frac{2\pi^2}{3}
\nonumber \\* &\qquad\qquad\;\;
            - \frac{2x \, (1+x) \, (1-2x)}{(1-x)^2 \, (1+2x)} \, \ln x
            + \frac{5 + 9x - 6x^2}{2(1-x) \, (1+2x)} \,
          \biggr]
  \,,
\end{align}
which agrees with the first-order
QCD correction to the top decay width neglecting the bottom mass, first 
obtained in~\cite{Jezabek:1988iv}. 
$\delta \Gamma_t^{(1)}$ is given by the sum of the hard-region contributions
from diagrams $h_{1c}$, $h_{1f}$, $g_1$, $g_2$, $g_6$ and from one half of
$h_{1d}+h_{1e}$. The hard-region contribution from diagrams $h_{1d}$ and
$h_{1e}$ is actually equal to the top-field renormalization contribution
from the counterterms of both vertices adjacent to the top line ($2 \times
\delta_t/2$ multiplied with the NLO diagram~$h_1$).
The above-mentioned set of diagrams provides virtual and real corrections
which only affect the upper top line. In the on-shell top case which is
effectively taken by the leading hard-region contribution, these
corrections precisely correspond to the ${\cal O}(\alpha_s)$ corrections to
the on-shell decay process $t\to b W^+$.
In the result~(\ref{eq:sigmatot}), $\delta \Gamma_t^{(1)}$ multiplies
exactly the endpoint-divergent $\Gamma_t^{\rm Born}/\Lambda$ term
in the NLO non-resonant result of~\cite{Beneke:2010mp}, such that $\delta
\Gamma_t^{(1)}$ may be dropped from our NNLO result if the top width in
the NLO result is replaced by its $\alpha_s$-corrected version
$\Gamma_t^{\rm Born} (1 + \delta\Gamma_t^{(1)})$.

Eq.~(\ref{eq:sigmatot}) is the main result of this work. 
As mentioned before, all ultraviolet ($1/\eps_\UV$) and infrared
($1/\eps_\IR$) singularities are canceled (independently of each other) in
this total contribution. The only remaining poles are found at order $(\Lambda/m_t)^0$; they
represent $1/\eps_\ep$ endpoint singularities and originate purely from the potential region.
When the $1/\eps$ finite-width divergences~(\ref{divsigmatotal})
from the NNLO resonant contributions in NRQCD are added to 
the NNLO non-resonant result (\ref{eq:sigmatot}), we observe a total 
cancellation of the $1/\eps$ divergences.\footnote{We have to set $s=4m_t^2$ and 
$M_Z^2=m_t^2 x/ c_w^2$, as done in the calculation of the absorptive matching coefficients
$C_p^{(v/a),\rm abs}$ \cite{Hoang:2004tg}, in order  to find an exact cancellation.} 
For the cancellation of the dependence on the renormalization scale
$\mu_{\rm soft}\sim m_t v$ associated 
with the finite-width divergences we have replaced 
consistently
\begin{equation}
 \bigg( \frac{\mu^2}{m_t^2} \bigg)^{3\eps} \rightarrow  \;
 \bigg( \frac{\mu^2_{\rm hard}}{m_t^2} \biggr)^{\eps} \bigg( \frac{\mu_{\rm soft}^2}{m_t^2} \bigg)^{2\eps}
\end{equation}
in the NNLO non-resonant amplitudes and taken $\mu_{\rm hard}=m_t$.
This is justified by the fact that our
$1/\eps_\ep$ singularities originate from one hard integration (associated
with the decay $t \to b W^+$) and two integrations with contributions from
a smaller scale $\mu_{\rm soft} \ll \mu_{\rm hard}$, stemming from the
on-shell limit of the top quarks and from potential-region gluon momenta.
Also in the resonant NNLO amplitude, one of the subgraphs always
corresponds to a hard contribution, {\it i.e.} to a matching coefficient
between QCD and NRQCD which is evaluated with $\mu_{\rm hard}=m_t$ (and in
$d=4$ dimensions). By setting consistently $\mu_{\rm hard}=m_t$ in our
contributions, too, the explicit dependence on $\mu_{\rm soft}$ associated
with the top-quark instability in the NNLO resonant contributions
({\it i.e.} the $\log \mu_{\rm soft}$ terms proportional to $\Gamma_t$ or
$C_p^{(v/a),\rm abs}$) cancels against the $\mu_{\rm soft}$-dependence of
the NNLO non-resonant endpoint-singular contributions~(\ref{eq:sigmatot}),
together with the cancellation of the $1/\eps$ poles. The remaining
logarithms of kinematic scales in this combination of resonant and
non-resonant pieces for the inclusive $t\bar{t}$ cross section with $bW$
invariant-mass cuts are of the form $\log (m_t v/\Lambda)$, in addition to
logarithms of the velocity in the resonant EFT matrix elements.

\section{Comparison with other approaches}
\label{sec:comparison}

\subsection{Comparison with the phase-space matching approach}
\label{sec:psm}

Our result (\ref{eq:sigmatot}) for the non-resonant contributions to the total cross section
with an invariant-mass cut $\Lambda^2$ on the $bW$ subsystems was obtained 
for the case of loose cuts, $\Lambda^2 \gg m_t E\sim m_t\Gamma_t$.
The condition of loose cuts
implies that the invariant-mass constraint only enters in the 
computation of the non-resonant contributions, while 
matrix elements on the resonant side are unaffected, as explained
in Refs.~\cite{Beneke:2010mp,Actis:2008rb}.
The endpoint-divergent terms presented here are equivalent to the first
terms in the expansion in $\Lambda/m_t$ of the full NNLO non-resonant
result.

Interestingly enough,
an alternative approach~\cite{Hoang:2010gu} (named {\it phase-space matching})
has determined the matching coefficients
of the four-electron operators providing the non-resonant 
contributions to the cross section with invariant-mass cuts on the $bW$ pairs
in the same range, $m_t\Gamma_t\ll \Lambda^2 \ll m_t^2$,
but through calculations involving only the matrix elements in the 
non-relativistic effective theory. In the phase-space matching (PSM)
approach, symmetric cuts on the invariant masses of the top and antitop
decay products restrict the integration over the top and antitop 
momenta in the resonant diagrams contributing to~(\ref{eq:Ares}).
Then the resulting
cut integrals are expanded assuming that 
$\Lambda^2 \gg m_t \Gamma_t$, but still smaller than $m_t^2$
so that the non-relativistic expansion carried out by NRQCD is still valid, 
yielding 
\begin{equation}
\label{eq:psm}
C(\alpha_s)\times \frac{\Gamma_t}{\Lambda} \times \sum_{n,m,k=0} \bigg[ \biggl(\frac{m_t \Gamma_t}{\Lambda^2}\biggr)^n 
\times \bigg(\frac{\Lambda^2}{m_t^2}\bigg)^m \bigg]\bigg(\alpha_s \frac{m_t}{\Lambda}\bigg)^k
+ \, \sigma_{\rm NRQCD}(\infty)
  \,,
\end{equation}
where $\sigma_{\rm NRQCD}(\infty)$ is the NRQCD resonant cross section without invariant-mass 
restrictions (equivalently, for $\Lambda\to \infty$).
The first term in~(\ref{eq:psm}) with powers of $\Lambda$
can be understood as minus the contributions to the NRQCD
matrix elements from the regions where the invariant-mass constraints are not fulfilled,
which are then subtracted from the unrestricted cross section by means of (\ref{eq:psm}).
The powers of $\Lambda^2/m_t^2$ in (\ref{eq:psm}) arise from cut diagrams with 
relativistic corrections, while factors of $\alpha_s m_t/\Lambda$ are
introduced by  cut diagrams with Coulomb-like gluons. $C(\alpha_s)=C_0+C_1 \alpha_s+\dots$ 
is a hard coefficient coming, for instance, from the matching
of the vector current that produces the non-relativistic $t\bar{t}$.
Assuming for the power-counting that $\Lambda\sim {\cal O}(m_t)$, the NLO terms
in the phase-space matching approach correspond to the terms $n=k=0$ in (\ref{eq:psm}) 
with $C(\alpha_s)\simeq C_0$,
while at NNLO we have to retain those with $n=0$ and $k=1$ plus the ${\cal O}(\alpha_s$)
correction to $C(\alpha_s)$ times the NLO term.
The whole computation is equivalent 
to the non-relativistic expansion of the full-theory squared matrix 
elements containing the double-resonant diagrams for 
$e^+e^-\to t\bar{t}\to W^+W^-b\bar{b}$ and their interference with 
the diagrams for $e^+e^-\to W^+W^- b\bar{b}$ having only either the 
top or the antitop in intermediate stages. 
However, the full-theory contributions coming from the square of single-top and 
pure background diagrams (the so-called remainder contributions 
in~\cite{Hoang:2010gu}) cannot be reproduced by EFT resonant diagrams,
and thus have to be computed with external tools in the phase-space 
matching approach. For moderate invariant-mass cuts it was shown 
in~\cite{Hoang:2010gu} through a numerical comparison with the 
full-theory tree-level $e^+e^-\to W^+W^- b\bar{b}$
cross section that the remainder contributions are small at NLO.
Indeed, in the computation of non-resonant corrections at this order within the 
unstable-particle EFT ({\it i.e.} the $n=k=0$ terms in (\ref{eq:psm})),
the remainder contributions are reproduced
by diagrams $h_{5}\,$--$\,h_{10}$ (see Fig.~\ref{fig1})
that contribute first with $\Gamma_t \Lambda^3/m_t^4$ terms in the
$\Lambda/m_t$-expansion, corresponding to $m=2$ in (\ref{eq:psm}).
For the first two NLO terms 
in the series of (\ref{eq:psm}), $m=0,1$, it was checked in~\cite{Beneke:2010mp} 
that they agree with the corresponding terms in the expansion of the
full NLO non-resonant result. 

Under the assumption that the QCD corrections to the remainder contributions
are also small in the above-mentioned range of invariant-mass cuts, 
the phase-space matching contributions at ${\cal O}(\alpha_s)$
were also computed in~\cite{Hoang:2010gu}. We can now compare the NNLO terms of the
phase-space matching series with those computed in this work, Eq.~(\ref{eq:sigmatot}):
the same coefficients in the $m_t^2/\Lambda^2$,  $m_t/\Lambda$ and 
$(m_t/\Lambda)^0 \log \Lambda$ contributions are found from that
comparison.\footnote{We note that 
in the phase-space matching approach, the term $m_t/\Lambda \times \delta\Gamma^{(1)}_t$
in~(\ref{eq:sigmatot}) is generated by including in the NRQCD top-quark propagators 
the ${\cal O}(\alpha_s)$ corrections to the top width, 
which was not done explicitly in the analysis of~\cite{Hoang:2010gu} because
$\Gamma_t$ was considered an input parameter there. The rest of the $m_t/\Lambda$ correction in~(\ref{eq:sigmatot})
arises in the phase-space matching approach from the product
of the ${\cal O}(\alpha_s)$ correction to the vector-current matching coefficient 
($-2\CF\,\alpha_s/\pi$) at both $t\bar{t}$-vertices with the NLO phase-space 
matching contributions.}

A simple argument can be used to explain why the two methods yield
the same series expansion in $(\Lambda/m_t)$ of the 
non-resonant contributions, despite the different starting points in the respective calculations. 
In our method, the full-theory amplitude is expanded first assuming that top-quark
lines are off-shell, {\it i.e.} $|p_t^2-m_t^2|\sim m_t^2$, which
drops the top self-energy corrections
from the full-theory top-quark
propagator, and then the $(\Lambda/m_t)$ series arises from further 
expanding the resulting amplitude around the on-shell limit, \textit{i.e.} for $|p_t^2-m_t^2| \ll m_t^2$.
The last expansion is equivalent to considering that the momentum $p_t$ is potential,
since the antitop on-shell condition sets $p_t^0-m_t=(p_t^2-m_t^2)/(4m_t) \sim
\vec{p}_t^{\;2}/m_t \ll m_t$.
On the other side, the PSM computation starts from 
the full-theory amplitude expanded for
nearly on-shell (potential) top quarks, with $p_t^0-m_t\sim \vec{p_t}^2/m_t\sim m_t v^2$, 
producing $\vec{p}_t^{\;2}/m_t^2$ and $(p_t^0-m_t)/m_t$ corrections. Then the latter are 
transformed into $(\Lambda^2/m_t^2)$
terms from hard-$p_t$ momentum regions ($|p_t^2-m_t^2|\sim m_t^2$) where
we can expand the integrand taking the off-shellness of $p_t$ much larger
than the non-relativistic scales $\Gamma_t$ and $E$.
The first important observation is that in both methods a double expansion 
of the integrand in $p_t$ (according to hard and potential region scalings) is performed, 
but in reversed order.
In the method of regions, such double expansions are known as overlap
contributions~\cite{Jantzen:2011nz}. 
Adopting the notation of the latter reference, 
we can denote as $T^{\rm (p)}T^{\rm (h)}I$ the integrand resulting from expanding the full-theory
integrand~$I$ (including the Dirac delta functions from the Cutkosky rules)
according to first hard and then potential momentum, 
and $T^{\rm (h)}T^{\rm (p)}I$ the same expansion taken in the reversed order, 
as effectively done in the PSM approach.
Within the expansion by regions, especially in simple cases with only two
relevant regions (here hard and potential), double expansions usually yield
the same doubly-expanded integrand, whether the two expansions are executed
in one order or the other. So we can expect that the identity
$T^{\rm (p)}T^{\rm (h)}I = T^{\rm (h)}T^{\rm (p)}I \equiv T^{\rm (h,p)}I$ holds.
Let us consider in the 
following at first the case without invariant-mass restrictions, so that $I$ does not 
depend on any cut~$\Lambda$. Then our second important remark is that
the integration of the integrand $T^{\rm (h,p)}I$ over $p_t$
vanishes in dimensional regularization, 
\begin{align}
\label{eq:Fhp}
F^{\rm (h,p)} &= \int d^d p_t \,  T^{\rm (h,p)}I = 0
  \,,
\end{align}
because after doubly expanding the integrand according to hard and potential momenta, there is 
no scale left in the integrand (recall that we have not imposed any invariant-mass constraints
in $I$, so the integration limits are unbounded). If we now rewrite (\ref{eq:Fhp}) using
$1=\theta(x-y)+\theta(y-x)$ as
\begin{align}
\label{eq:cancel}
0 = \int d^d p_t \, \theta(\Lambda^2 - (m_t^2- p_t^2)) \, T^{\rm (h,p)}I
    + \int d^d p_t \, \theta((m_t^2-p_t^2)-\Lambda^2) \, T^{\rm (h,p)}I
  \,,
\end{align}
we readily identify the first term 
as the Taylor series in $(\Lambda/m_t)$ of the non-resonant contributions
computed in our approach for an invariant-mass cut of the form (\ref{eq:DeltaM}). 
The second term corresponds
to the resonant NRQCD amplitude further expanded assuming hard~$p_t$ and integrated
outside the region of $p_t^2$ allowed by the invariant-mass cuts. That is precisely
the quantity that yields minus the series in the first term of~(\ref{eq:psm})
with powers of~$\Lambda$, obtained
in the phase-space matching approach. Therefore, from (\ref{eq:cancel}) we conclude that
the PSM contributions must be equal to the series expansion obtained with
the unstable-particle EFT, as we have shown by explicit computation up to 
terms of order $(m_t/\Lambda)^0 \log \Lambda$. The equivalence between both series is lost at the order
where terms coming from full-theory diagrams with just one top or antitop, or 
with no tops at all, which are not described by NRQCD, first contribute.
As mentioned before, these terms arise at NLO from diagrams $h_5\,$--$\,h_{10}$, and are
of order $\Gamma_t\Lambda^3/m_t^4$. Since the ${\cal O}(\alpha_s)$
corrections to diagrams $h_5\,$--$\,h_{10}$ do not introduce negative powers of $(1-t)$,
we also expect that the NNLO non-resonant contributions in the phase-space matching series 
start to differ from ours at order $\alpha_s \, \Gamma_t\Lambda^3/m_t^4$.

The reasoning above also provides further insight on the regions (of gluon
momenta) which contribute to the endpoint-singular NNLO corrections.
In scaleless overlap contributions like~(\ref{eq:Fhp}) where the integrand
is doubly expanded according to both regions, singularities from domains of
hard and potential top momenta~$p_t$ cancel each other to yield zero.
Exactly the same cancellation of singularities happens between the
contributions of the individual (hard or potential) regions, originating
from the respective integration domain where the top momentum approaches
the scaling of the opposite region:
The ultraviolet singularities (from the
hard-$p_t$ limit) present in the integrals of the resonant contribution
(which have been expanded for potential~$p_t$) are canceled by the endpoint
singularities (from the potential-$p_t$ limit) of the non-resonant
contribution (whose integrals have been expanded for hard~$p_t$).
This cancellation occurs individually for the different scalings of gluon
momenta. It is known that the $1/\eps$ finite-width divergences on the
resonant side originate purely from potential (Coulomb) gluons. So their
cancellation must be provided purely by potential gluon momenta in the
endpoint-divergent non-resonant diagrams. This is exactly what we have
found in our calculation for the origin of the $1/\eps_\ep$ terms from
potential-region contributions.

Similarly, it is known that ultrasoft gluons do not contribute to the total
resonant NRQCD amplitudes at NNLO. So it comes without surprise that we
observe a cancellation of the ultrasoft-region contributions to the
endpoint-divergent non-resonant corrections (see the remarks at the end of
Sec.~\ref{subsec:realgluon}).

\subsection{Comparison with an approach based on expanding in
  \boldmath$\rho = 1-M_W/m_t$}
\label{sec:rho}

A different path has been taken by the authors of Ref.~\cite{Penin:2011gg} to provide an estimate of the
NNLO non-resonant corrections. They have expanded the NNLO non-resonant contributions to the $t\bar{t}$ total
cross section in powers of $\sqrt{\rho}$, where $\rho=1-M_W/m_t\approx
0.5$, and calculated the first two terms in this expansion.
For the NLO non-resonant corrections, the
leading-order term in $\rho$ has been shown to deviate from our exact result~\cite{Beneke:2010mp} by less than 5\%,
despite the fact that the approximation for the individual diagrams is much less accurate. Such a numerical comparison
is not possible at NNLO where the exact result is not known, and our approximation, Eq.~(\ref{eq:sigmatot}), does not
correspond to the total cross section, but to the cross section with invariant-mass cuts applied in the $bW$ pairs.

We notice though important differences between the approach followed in~\cite{Penin:2011gg} and ours for
the NNLO result.
According to the authors of~\cite{Penin:2011gg} the leading-order term in $\rho$ of the NNLO 
non-resonant contributions arises from diagram $h_{1a}$ and
contains an energy-dependent term of the form $\log (v/\rho)$. The latter is generated in their approach
because the scale $m_t v^2$ in the
top-quark propagators is kept as an infrared regulator for the (infrared-divergent) loop integration of the non-resonant
amplitude, despite the fact that the integral is saturated by the region $|\vec{p}|\sim \rho^{1/2}m_t \gg m_t v$. 
Let us remark that in our approach logarithms of the velocity can only appear in the resonant matrix elements,
because the hard-momentum expansion performed in the non-resonant diagrams  
drops terms of order $E,\,\Gamma_t \sim m_t v^2$ from the top-quark propagators and, as a consequence, 
an energy-dependence there only arises beyond NNLO as powers of $E$. 
The infrared divergence in~\cite{Penin:2011gg} is analogous to the $1/\eps_\ep$ endpoint divergences we find within the 
unstable-particle EFT formalism, and must be compensated by an ultraviolet divergence arising from a resonant contribution
in order to render the leading-order result in the $\rho$-expansion independent of the regularization scheme.%
\footnote{The leading-order
NNLO non-resonant term in~\cite{Penin:2011gg} is proportional to $\alpha_s \Gamma_t/\rho$.
While the NNLO relativistic corrections to the resonant diagram with a Coulomb potential, Fig.~\ref{figCoulomb},
only give terms scaling as $\alpha_s \Gamma_t \times {\cal O}(\rho^0)$, the corrections arising
from the insertion of the (electroweak) absorptive matching coefficients $C_p^{(v/a),\rm abs}$ 
into Fig.~\ref{figCoulomb} yield terms scaling also as $\alpha_s \Gamma_t/\rho$. Part of the latter provides
the counterpart for the infrared divergence in the leading-order result from~\cite{Penin:2011gg}.}
The need for an ultraviolet counterpart for the result of~\cite{Penin:2011gg}
becomes even clearer if the
infrared divergence on the non-resonant side is regulated dimensionally, which substitutes the $\log v$ above by
a $\log \mu$ plus (potentially) additional finite terms, thus introducing an explicit scheme-dependence 
through the renormalization scale~$\mu$. We can easily do this exercise by taking the 
integral form of the leading-order term provided in Eq.~(4.1) of~\cite{Penin:2011gg} and performing 
it in $d$~dimensions. With the 
appropriate normalization factor for the cross section, this contribution reads
\begin{align}
  \label{eq:sigmaNNLOPenin}
  &\delta\sigma_{\rm non-res}^{(2),\,\rho^{-1}}
    = \frac{4\pi  \alpha^2}{3s} \,
      \Bigl[ Q_t^2 Q_e^2 
        + \frac{2 Q_e Q_t v_e v_t}{1-M_Z^2/(4m_t^2)} + \frac{(a_e^2+v_e^2)v_t^2}{(1-M_Z^2/(4m_t^2))^2} \Bigr] \,
       \frac{\Nc \Gamma_t^{\rm Born}}{m_t} \, \delta_{1a}^{(1)} \Big|_{\rho^{-1}}
  \,,
\end{align}
with 
\begin{align}
  \label{eq:delta1a1}
   \delta_{1a}^{(1)} \Big|_{\rho^{-1}} = 3\left[ \frac{1}{2\eps}+\ln\left(\frac{\mu^2}{\rho m_t^2}\right)+1\right]
   \frac{\CF \alpha_s}{\rho}
  \,.
\end{align}
As expected, the coefficient of the infrared-divergent $1/\eps$ term in $\delta\sigma_{\rm non-res}^{(2),\,\rho^{-1}}$
above agrees with the $1/\eps_\ep$ term 
which arises from our result for diagram $h_{1a}$, Eq.~(\ref{eq:H1a}) plus symmetric contributions, 
expanded at the leading order for small $\rho$ 
(note that $x=M_W^2/m_t^2=(1-\rho)^2$ and that the coefficient of
$v_t^\tL v_t^\tR/\eps_\ep$  in~(\ref{eq:H1a})
is proportional to $1/(1-x)\simeq 1/(2\rho)$). A quick inspection of the rest of the $1/\eps_\ep$
endpoint-divergent results computed in Sec.~\ref{sec:results} reveals that diagram $h_{1b}$ also contains
an infrared $1/\eps_\ep$ divergence proportional to $\rho^{-1}$. It can also be checked that 
the sum of the $\rho^{-1}/\eps_\ep$ divergences from diagrams $h_{1a}$ and $h_{1b}$ 
(including symmetric contributions) cancels against 
the first term in the $\rho$-expansion of the finite-width divergence on the resonant side, Eq.~(\ref{divsigmatotal}). 
The contribution from diagram $h_{1b}$ has not been considered, however,
in the analysis of the $1/\rho$ term of~\cite{Penin:2011gg}. We therefore conclude 
that in its present form the leading-order NNLO non-resonant result of~\cite{Penin:2011gg}
is incomplete,  and cannot be consistently added as a correction to the
known NNLO contributions from the resonant side.

Finally, for implementing a loose cut on the $bW$ invariant masses in the
integrals for the NLO non-resonant contributions expanded in the
parameter~$\rho$, the authors of~\cite{Penin:2011gg} have provided the
replacement rule $\rho\,m_t^2\to \Lambda^2/2$ applied to the arguments of the
theta-functions of their integrals. Adopting the same rule for the ${\cal O}(\alpha_s)$
contribution that yields the leading non-resonant term at NNLO in the approach of~\cite{Penin:2011gg}, 
the same $m_t^2/\Lambda^2$ term as in our result~(\ref{eq:sigmatot}) is obtained. 
Our subleading terms of order $m_t/\Lambda$ cannot directly be compared
to~\cite{Penin:2011gg}.

\section{Final results}
\label{sec:finalresults}

In this final section we compare numerically
the first terms in the $(\Lambda/m_t)$-series 
of the NNLO non-resonant corrections with the NLO ones
as well as with the leading-order EFT approximation. This is done
for the $e^+e^-\to W^+W^-b\bar{b}$ cross section with invariant-mass
cuts in the $bW$ subsystems of the form (\ref{eq:DeltaM}), where $\Lambda^2=2m_t \Delta M_t-\Delta M_t^2$.  
The NNLO non-resonant terms are given by the endpoint-singular
contributions computed in Sec.~\ref{sec:results}, and have been collected
in Eq.~(\ref{eq:sigmatot}). The NLO non-resonant corrections, $\sigma_{\rm non-res}^{(1)}$,
are obtained from the results given in \cite{Beneke:2010mp}, which are too lengthy
to be reproduced here.
We present NLO and NNLO contributions separately in this section,
\textit{i.e.} ``NNLO'' always refers to the pure second-order corrections
without including the NLO result.

A dependence on the invariant-mass cut 
$\Delta M_t$ enters first at NLO through the non-resonant contributions for the case
of loose cuts ({\it i.e.} $\Delta M_t \gg \Gamma_t$). The leading-order
cross section $\sigma_{t\bar{t}}^{(0)}$ is given entirely by the leading-order resonant contribution
(\ref{eq:Ares}), which sums Coulomb corrections proportional to $(\alpha_s/v)^n$ to all 
orders in the strong coupling. Its analytic expression, following the same conventions
as in this paper, can be found in \cite{Beneke:2010mp}. $\sigma_{t\bar{t}}^{(0)}$
depends on the renormalization scale $\mu_{\rm soft}$ only through $\alpha_s(\mu_{\rm soft})$.
\begin{figure}[t]
\begin{center}
\includegraphics[width=0.8\textwidth]{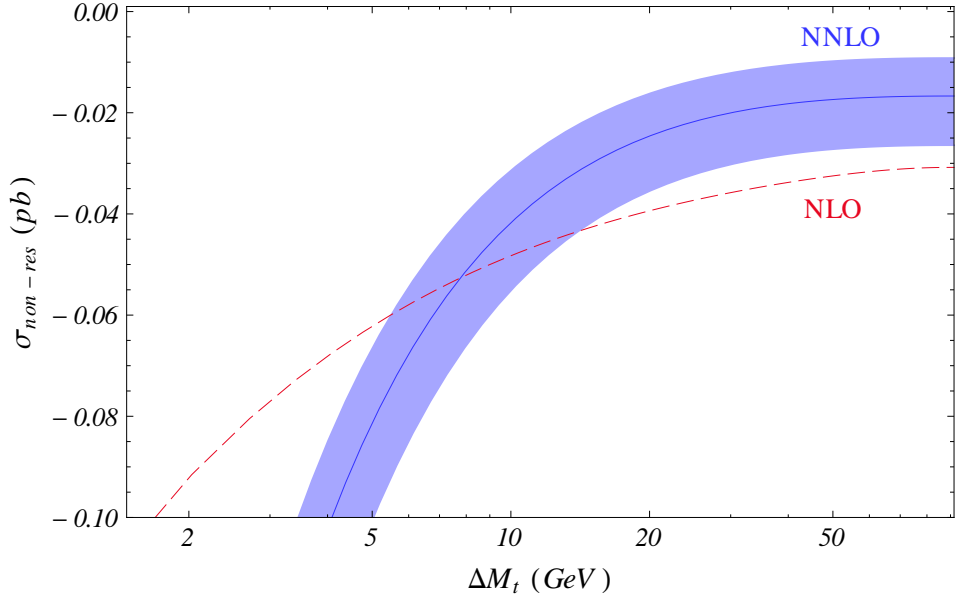}

\caption{Endpoint-singular NNLO non-resonant contribution to the $t\bar{t}$
cross section computed at $s=4m_t^2$ as a function 
of the invariant-mass cut $\Delta M_t=m_t(1- \sqrt{1 - \Lambda^2/m_t^2})$.
The solid (blue) line corresponds to $\sigma_{\rm non-res}^{(2),\ep}$ in (\ref{eq:sigmatot})
dropping the $1/\eps_\ep$ term, and  
using $\alpha_s\equiv \alpha_s(\mu_{\rm soft})$ with $\mu_{\rm soft}=30~{\rm GeV}$.
The shaded (blue) band shows the same result varying 
$\mu_{\rm soft}$ in the interval 15--60~GeV. For comparison, the NLO non-resonant contribution
(dashed red line), $\sigma_{\rm non-res}^{(1)}$ from \cite{Beneke:2010mp},
is also shown.} 
\label{fig4}
\end{center}
\end{figure}

Figure~\ref{fig4} shows the contribution to the cross section from the NNLO
endpoint-singular terms as a function of the invariant-mass cutoff exactly at 
the threshold ($s=4m_t^2$).
The NNLO contribution corresponds to $\sigma_{\rm non-res}^{(2),\ep}$ in (\ref{eq:sigmatot})
with the endpoint divergence $1/\eps_\ep$ removed. It uses
$\alpha_s\equiv\alpha_s(\mu_{\rm soft})$ plus the Standard-Model input parameters
\begin{equation}
M_Z = 91.1876\,{\rm GeV} \;,\qquad M_W = 80.398\,{\rm GeV}\;,\qquad m_t =172.0\,{\rm GeV} \,,
\nonumber
\end{equation}
\begin{equation}
G_\mu =1.166367\times 10^{-5}\,{\rm GeV}{}^{-2} \;,\qquad
V_{tb}=1 \,,
\label{eq:input}
\end{equation}
whereas the on-shell Weinberg angle $\cw=M_W/M_Z$ and the fine-structure 
constant in the $G_\mu$-scheme, $\alpha\equiv \sqrt{2}G_\mu M_W^2 \sw^2/\pi$, 
are derived quantities.
The solid (blue) line is obtained for $\mu_{\rm soft}=30$~GeV, where
$\alpha_s(30\,\text{GeV}) = 0.142$, whereas 
the shaded band displays the effect of varying the scale 
$\mu_{\rm soft}$ from 15 to 60~GeV (lower values of the scale corresponding
to more negative contributions). The dashed (red) line is the 
NLO non-resonant contribution $\sigma_{\rm non-res}^{(1)}$ from \cite{Beneke:2010mp}
which is also shown for comparison. Both the NLO and NNLO non-resonant
corrections give a negative shift.

The values shown in Fig.~\ref{fig4}
range from $\Delta M_t=\Gamma_t\simeq 1.46$~GeV up to the maximum value allowed 
by the kinematics,  $\Delta M_{t,\rm max}=m_t-M_W\simeq 91.6$~GeV, which corresponds
to the total cross section. Recall that the NNLO non-resonant terms computed in this work
are a valid description for moderate invariant-mass cuts, satisfying
$\Gamma_t\ll \Delta M_t \ll m_t$. For tight cuts ($\Delta M_t\lesssim
\Gamma_t$), which are not studied here, the expansion
by regions dictates that the dependence on $\Delta M_t$ is taken into account in the 
resonant part of the amplitude, while the non-resonant contributions are absent in this case (see~\cite{Beneke:2010mp}).
The further requirement $\Delta M_t\ll m_t$ is a consequence 
of the expansion around the endpoint that produces the result (\ref{eq:sigmatot})
for the non-resonant contributions. {}From Fig.~\ref{fig4} we observe that in a
moderate $\Delta M_t$-range ($\Delta M_t \gtrsim 6\,\Gamma_t \approx
9\,\text{GeV}$) the NNLO non-resonant corrections are always 
smaller (in absolute value) than the NLO ones, and (because of the higher
singularity in $\Lambda$ or $\Delta M_t$) they become more 
negative when the available phase space for the $bW$ pairs gets restricted
by tightening the invariant-mass cut. The ratio between the NNLO and NLO 
non-resonant contributions
ranges approximately from 0.9 to 0.5 for  $\Delta M_t$ in the interval $(10,40)$~GeV.

\begin{figure}[t]
\begin{center}
\includegraphics[width=0.8\textwidth]{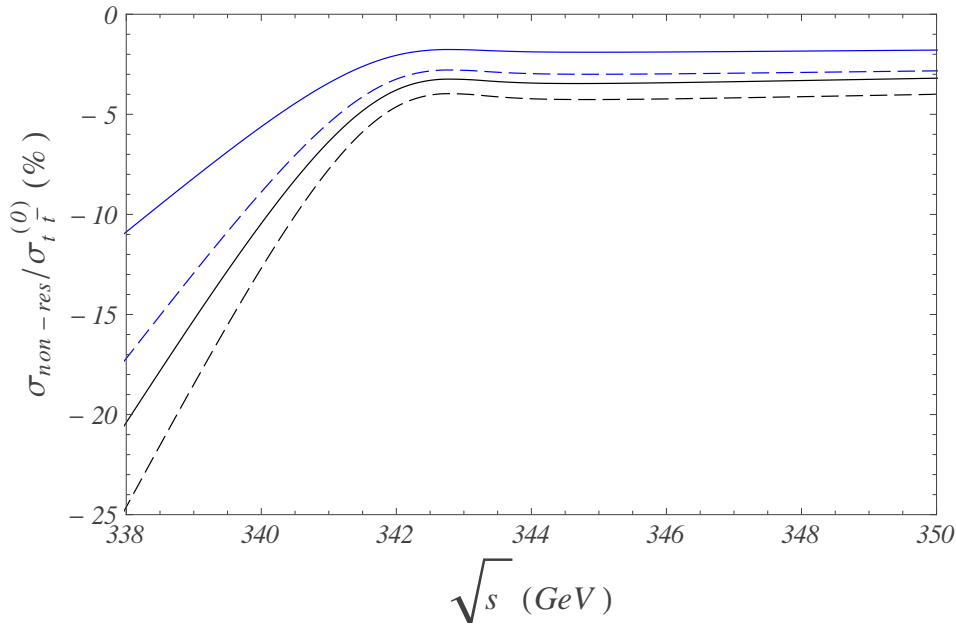}

\caption{Relative sizes of the non-resonant corrections with respect to
the $t\bar{t}$ LO cross section in percent: $\sigma_{\rm non-res}^{(2),\ep}/\sigma_{t\bar{t}}^{(0)}$
(upper blue lines) and $\sigma_{\rm non-res}^{(1)}/\sigma_{t\bar{t}}^{(0)}$ (lower black lines, from
\cite{Beneke:2010mp}). Solid (dashed) lines correspond to an invariant-mass cut $\Delta M_t=35$~GeV
($\Delta M_t=15$~GeV). The renormalization scale in the NNLO non-resonant contribution
has been set to $\mu_{\rm soft}=30$~GeV, and we have chosen $\alpha_s(30~{\rm GeV})=0.142$ for the value
of the QCD coupling that enters also in the LO result.}
\label{fig5}
\end{center}
\end{figure}

The relative size of the endpoint-singular NNLO corrections with respect to
the LO cross section as a function of the centre-of-mass energy in the
threshold region
is displayed in Figure~\ref{fig5} for two different values of the 
invariant-mass cut, $\Delta M_t=35$~GeV (upper blue solid line) and
$\Delta M_t=15$~GeV (upper blue dashed line).
The corresponding curves 
for the NLO result are also shown (lower black lines).  
At threshold energies, $|\sqrt{s}-2m_t|\ll 2m_t$, the non-resonant corrections are
almost energy-independent, with a mild linear energy-dependence introduced from
the $Z$ and photon propagators. 
The NNLO non-resonant
corrections for the chosen values of $\Delta M_t$ give a constant negative 
shift of about $2-3\%$ above the threshold where the LO cross section is also constant. 
Below the peak region, where the LO result vanishes rapidly, the relative size of the 
non-resonant corrections is very large, up to 10\% for $\Delta M_t=35$~GeV
in the energy range shown in Fig.~\ref{fig5}.
In absolute value, the endpoint-singular NNLO non-resonant corrections 
amount to \mbox{32--28~fb} for $\Delta M_t=15$~GeV and \mbox{20--18~fb} for $\Delta M_t=35$~GeV
when the centre-of-mass energy is varied within the interval $(338,350)$~GeV.

For an analysis of the impact of higher-order non-resonant corrections for moderate invariant-mass cuts 
not considered in the present work, namely the non-logarithmic terms of order $(\Lambda/m_t)^0$ 
and a large part of the N${}^3$LO contributions, we refer
the reader to~\cite{Hoang:2010gu}.

\section{Summary}
\label{sec:conclusion}

The corrections induced by off-shell top-quark decay and by other non-resonant production 
processes of the physical $W^+W^- b\bar{b}$ final state 
are a missing piece towards the prediction of the threshold top-quark pair production cross section 
at the third order (NNLO). Phenomenologically, they are needed to match the accuracy
of the well-known QCD corrections to the resonant production, 
as required for the precision attainable in the top-mass
determination at a future $e^+e^-$ collider.
The computation of the full set of NNLO non-resonant corrections 
represents a non-trivial task,
involving ${\cal O}(100)$ diagrams that include 1-loop virtual-QCD as well as
tree-level gluon-radiation corrections to the $bWt$ final state. 

At the theory level, the NNLO non-resonant corrections are also mandatory,
because the purely resonant cross section contains at the same order an uncanceled 
ultraviolet divergence,
${\rm div} \, \sigma_{\rm res}^{\mathrm{NNLO}} \propto \alpha_s \Gamma_t/\epsilon$~(\ref{divsigmatotal}),
which must be compensated by a divergence with opposite sign on the non-resonant side 
in order to yield
a regularization-independent result for this observable.
In this work we have identified the divergences in the NNLO non-resonant amplitudes
which provide such a cancellation. They originate at the endpoint of the phase-space
integration over the $bW$ invariant mass in virtual diagrams with
$bWt$ final states.  We have extracted these endpoint divergences through an expansion 
of the relevant NNLO non-resonant diagrams around the endpoint. The
expanded integrals involve the scale $\Lambda^2$ of the invariant-mass cut
in the $bW$ system. In this way, apart from
the $1/\eps$ divergences, we obtain the
endpoint-singular terms $(m_t/\Lambda)^2$, $(m_t/\Lambda)$ and $(m_t/\Lambda)^0\log \Lambda$
which correspond to the first terms in the expansion in powers of $(\Lambda/m_t)$.
This series provides a rigorous approximation of the NNLO non-resonant 
contributions
to the $e^+e^-\to W^+W^- b\bar{b}$ cross section with symmetric
invariant-mass cuts of size $\Lambda^2\approx 2m_t\Delta M_t $ applied
to the $bW$ pairs, as long as $\Delta M_t$ is much smaller than the top mass but
significantly larger than the top width (equivalently, if $m_t \Gamma_t \ll
\Lambda^2 \ll m_t^2$).

Our analytic result agrees with the one obtained for the same observable within the 
phase-space matching approach~\cite{Hoang:2010gu}. On the other hand, by comparing
the infrared structure of our result with the one obtained in~\cite{Penin:2011gg} 
at leading order in the expansion for small $\rho=1-M_W/m_t$,
we conclude that the latter misses one contribution
and that it cannot be combined with the NNLO resonant corrections in 
a regularization-scheme independent way. 
Numerically we find that within the above range of invariant-mass cuts the NNLO non-resonant
contributions produce a negative shift of about $2-3\%$ with respect to the leading-order
$t\bar{t}$ cross section above threshold. For energies below the peak, where non-resonant
production is known to dominate over the (subleading) resonant terms, the corrections
reach up to $10-15\%$.

The presence of endpoint singularities in the NNLO non-resonant contributions represents
an additional complication for their calculation, since these singularities have
to be subtracted  from the amplitude, together with the standard
soft-collinear divergences due to gluon radiation.
The analysis performed in this work, which identifies and evaluates the endpoint-singular terms, 
thus provides a necessary step towards the computation of the full set 
of NNLO non-resonant contributions to $e^+e^- \to W^+W^- b\bar{b}$ and, consequently,
towards having a complete NNLO theoretical prediction for $t\bar{t}$ production  near threshold 
that accounts for the effects related to top-quark decay in a consistent manner. 

\vspace*{0.2cm}
\noindent
\subsubsection*{Acknowledgments}
We thank M.~Beneke for helpful discussions and for his comments on the manuscript. 
We also thank A.~A.~Penin for discussions on the comparison with their result.
This work is supported by the DFG Sonder\-forschungsbereich/Transregio~9
``Computergest\"utzte Theoretische Teilchenphysik''.
The work of P.~R. is partially supported by MEC (Spain) under grants FPA2007-60323 and FPA2011-23778 and by the
Spanish Consolider-Ingenio 2010 Programme CPAN (CSD2007-00042).
Feynman diagrams have been drawn with the packages 
{\sc Axodraw}~\cite{Vermaseren:1994je} and 
{\sc Jaxo\-draw}~\cite{Binosi:2008ig}. We acknowledge the use of 
the computer programs \textsc{FORM}~\cite{Vermaseren:2000nd} and
\textsc{FeynCalc}~\cite{Mertig:1990an} for 
some parts of the calculation.



\begin{thebibliography}{99}

\bibitem{Martinez:2002st}
M.~Martinez and R.~Miquel,
\newblock Eur. Phys. J. C {\bf 27} (2003) 49, hep-ph/0207315.


\bibitem{Seidel:2013sqa}
  K.~Seidel, F.~Simon, M.~Tesar and S.~Poss,
  Eur.\  Phys.\  J.\ C {\bf 73} (2013) 2530,
  arXiv:1303.3758 [hep-ex].

\bibitem{Aaltonen:2012ra}
  T.~Aaltonen {\it et al.}  [CDF and D0 Collaborations],
  Phys.\ Rev.\ D {\bf 86} (2012) 092003,
  arXiv:1207.1069 [hep-ex].

\bibitem{ATLAS:2012coa}
  ATLAS Collaboration,
  ATLAS-CONF-2012-095.

\bibitem{Bigi:1986jk}
  I.~I.~Y.~Bigi, Y.~L.~Dokshitzer, V.~A.~Khoze, J.~H.~K\"uhn and P.~M.~Zerwas,
  Phys.\ Lett.\ B {\bf 181} (1986) 157.


\bibitem{Fadin:1987wz}
V.~S. Fadin and V.~A. Khoze,
\newblock JETP Lett. {\bf 46} (1987) 525.

\bibitem{Fadin:1988fn}
V.~S. Fadin and V.~A. Khoze,
\newblock Sov. J. Nucl. Phys. {\bf 48} (1988) 309.

\bibitem{Strassler:1990nw}
M.~J. Strassler and M.~E. Peskin,
\newblock Phys. Rev. D {\bf 43} (1991) 1500.


\bibitem{Caswell:1985ui}
  W.~E.~Caswell and G.~P.~Lepage,
  Phys.\ Lett.\ B {\bf 167} (1986) 437.

\bibitem{Hoang:2000yr}
  A.~H.~Hoang {\it et al.},
  Eur.\ Phys.\ J.\ direct C {\bf 2} (2000) 1,
  hep-ph/0001286.


\bibitem{Beneke:2005hg}
M.~Beneke, Y.~Kiyo and K.~Schuller,
\newblock Nucl. Phys. B {\bf 714} (2005) 67, hep-ph/0501289.

\bibitem{Beneke:2008ec}
  M.~Beneke, Y.~Kiyo and K.~Schuller,
  PoS (RADCOR 2007) 051,
  arXiv:0801.3464 [hep-ph].

\bibitem{Beneke:2008cr}
  M.~Beneke and Y.~Kiyo,
  Phys.\ Lett.\  B {\bf 668} (2008) 143,
  arXiv:0804.4004 [hep-ph].

\bibitem{Pineda:1997bj}
  A.~Pineda and J.~Soto,
  Nucl.\ Phys.\ Proc.\ Suppl.\  {\bf 64} (1998) 428,
  hep-ph/9707481.

\bibitem{Beneke:1998jj}
  M.~Beneke,
  hep-ph/9806429.

\bibitem{Beneke:1999qg}
  M.~Beneke, A.~Signer and V.~A.~Smirnov,
  Phys.\ Lett.\ B {\bf 454} (1999) 137,
  hep-ph/9903260.

\bibitem{Brambilla:1999xf}
  N.~Brambilla, A.~Pineda, J.~Soto and A.~Vairo,
  Nucl.\ Phys.\ B {\bf 566} (2000) 275,
  hep-ph/9907240.


\bibitem{Luke:1999kz}
  M.~E.~Luke, A.~V.~Manohar and I.~Z.~Rothstein,
  Phys.\ Rev.\ D {\bf 61} (2000) 074025,
  hep-ph/9910209.

\bibitem{Manohar:1999xd}
  A.~V.~Manohar and I.~W.~Stewart,
  Phys.\ Rev.\ D {\bf 62} (2000) 014033,
  hep-ph/9912226.

\bibitem{Hoang:2002yy}
  A.~H.~Hoang and I.~W.~Stewart,
  Phys.\ Rev.\ D {\bf 67} (2003) 114020,
  hep-ph/0209340.

\bibitem{Hoang:2000ib}
A.~H. Hoang, A.~V. Manohar, I.~W. Stewart and T.~Teubner,
\newblock Phys. Rev. Lett. {\bf 86} (2001) 1951, hep-ph/0011254.

\bibitem{Hoang:2001mm}
A.~H. Hoang, A.~V. Manohar, I.~W. Stewart and T.~Teubner,
\newblock Phys. Rev. D {\bf 65} (2001) 014014, hep-ph/0107144.

\bibitem{Pineda:2006ri}
A.~Pineda and A.~Signer,
Nucl. Phys. B {\bf 762} (2007) 67, hep-ph/0607239.

\bibitem{Hoang:2011it}
  M.~Stahlhofen and A.~Hoang,
  PoS (RADCOR 2011) 025,
  arXiv:1111.4486 [hep-ph].

\bibitem{Beneke:2003xh}
M.~Beneke, A.~P. Chapovsky, A.~Signer and G.~Zanderighi,
\newblock Phys. Rev. Lett. {\bf 93} (2004) 011602, hep-ph/0312331.

\bibitem{Beneke:2004km}
M.~Beneke, A.~P. Chapovsky, A.~Signer and G.~Zanderighi,
\newblock Nucl. Phys. B {\bf 686} (2004) 205, hep-ph/0401002.

\bibitem{Beneke:2007zg}
  M.~Beneke, P.~Falgari, C.~Schwinn, A.~Signer and G.~Zanderighi,
  Nucl.\ Phys.\  B {\bf 792} (2008) 89,
  arXiv:0707.0773 [hep-ph].


\bibitem{Beneke:2010mp}
  M.~Beneke, B.~Jantzen and P.~Ruiz-Femen\'ia,
  Nucl.\ Phys.\ B {\bf 840} (2010) 186,
  arXiv:1004.2188 [hep-ph].


\bibitem{Hoang:2008ud}
  A.~H.~Hoang, C.~J.~Rei{\ss}er and P.~Ruiz-Femen\'ia,
  Nucl.\ Phys.\ Proc.\ Suppl.\  {\bf 186} (2009) 403, 
  arXiv:0810.2934 [hep-ph].

\bibitem{Hoang:2010gu}
  A.~H.~Hoang, C.~J.~Rei{\ss}er and P.~Ruiz-Femen\'ia,
  Phys.\ Rev.\  D {\bf 82} (2010) 014005,
  arXiv:1002.3223 [hep-ph].


\bibitem{Hoang:2004tg}
  A.~H.~Hoang and C.~J.~Rei{\ss}er,
  Phys.\ Rev.\  D {\bf 71} (2005) 074022, 
  hep-ph/0412258.

\bibitem{RuizFemenia:2012ma}
  P.~Ruiz-Femen\'ia,
  arXiv:1203.0934 [hep-ph].



\bibitem{Penin:2011gg}
  A.~A.~Penin and J.~H.~Piclum,
  JHEP {\bf 1201} (2012) 034,
  arXiv:1110.1970 [hep-ph].

\bibitem{Beneke:1997zp}
  M.~Beneke and V.~A.~Smirnov,
  Nucl.\ Phys.\  B {\bf 522} (1998) 321,
  hep-ph/9711391.

\bibitem{Smirnov:2002pj}
  V.~A.~Smirnov,
  \emph{Applied asymptotic expansions in momenta and masses},
  vol.~177 of {\em Springer Tracts in Modern Physics}, Springer, Germany
  (2002).

\bibitem{Jantzen:2011nz}
  B.~Jantzen,
  JHEP {\bf 1112} (2011) 076,
  arXiv:1111.2589 [hep-ph].


\bibitem{Jezabek:1988iv}
  M.~Jezabek and J.~H.~K\"uhn,
  Nucl.\ Phys.\  B {\bf 314} (1989) 1.

\bibitem{Actis:2008rb}
  S.~Actis, M.~Beneke, P.~Falgari and C.~Schwinn,
  Nucl.\ Phys.\  B {\bf 807} (2009) 1,
  arXiv:0807.0102 [hep-ph].

\bibitem{Vermaseren:1994je}
  J.~A.~M.~Vermaseren,
  Comput.\ Phys.\ Commun.\  {\bf 83} (1994) 45.

\bibitem{Binosi:2008ig}
  D.~Binosi, J.~Collins, C.~Kaufhold and L.~Theussl,
  Comput.\ Phys.\ Commun.\  {\bf 180} (2009) 1709,
  arXiv:0811.4113 [hep-ph].

\bibitem{Vermaseren:2000nd}
  J.~A.~M.~Vermaseren,
  math-ph/0010025.

\bibitem{Mertig:1990an}
  R.~Mertig, M.~B\"ohm and A.~Denner,
  Comput.\ Phys.\ Commun.\  {\bf 64} (1991) 345.


\end{thebibliography}
\end{document}